\documentclass[pdftex,twocolumn,epjc3]{svjour3}          

\RequirePackage[T1]{fontenc}

\smartqed  

\RequirePackage{graphicx}
\RequirePackage{amssymb}
\RequirePackage{amsmath}
\RequirePackage{mathptmx}      
\RequirePackage{flushend}
\RequirePackage[numbers,sort&compress]{natbib}
\RequirePackage[colorlinks,citecolor=blue,urlcolor=blue,linkcolor=blue]{hyperref}

\journalname{Eur. Phys. J. C}

\begin{document}

\title{U-spin symmetry energy and hyperon puzzle in $\Lambda$-admixed neutron stars}


\author{Hao-Song You\thanksref{e1,tdli,yzu}
        \and
        Ting-Lan Yu\thanksref{yzu}
        \and
        Cheng-Jun Xia\thanksref{yzu}
        \and
        Sophia Han\thanksref{tdli,spa,skldmp}
        \and
        Ren-Xin Xu\thanksref{pku,kiaa}
}

\thankstext{e1}{e-mail: sjtu1736142@sjtu.edu.cn}

\institute{{Tsung-Dao Lee Institute, Shanghai Jiao Tong University},
  {Shanghai} {201210},
  {China}\label{tdli}
          \and
          {Center for Gravitation and Cosmology, College of Physical Science and Technology, Yangzhou University},
  {Yangzhou} {225009},
  {China}\label{yzu}
          \and
          {School of Physics and Astronomy, Shanghai Jiao Tong University},
  {Shanghai} {200240},
  {China}\label{spa}
          \and
          {State Key Laboratory of Dark Matter Physics, Shanghai Jiao Tong University},
  {Shanghai} {201210},
  {China}\label{skldmp}
          \and
          {School of Physics and State Key Laboratory of Nuclear Physics and Technology, Peking University},
  {Beijing} {100871},
  {China}\label{pku}
          \and
          {Kavli Institute for Astronomy and Astrophysics, Peking University},
  {Beijing} {100871},
  {China}\label{kiaa}
}

\date{Received: date / Accepted: date}

\maketitle

\begin{abstract}
By combining the ($u$,$d$) I-spin doublets or ($d$,$s$) U-spin doublets, the SU(3) flavor symmetry of light quarks can be decomposed into SU(2)$_I\times$U(1)$_Y$ or SU(2)$_U\times$U(1)$_Q$ subgroups, which have been widely adopted to categorize hadrons and their decay properties. The standard I-spin counterpart for the interactions among nucleons, namely the nuclear symmetry energy $E_\mathrm{sym}(n_\mathrm{b})$, characterizes the variation of binding energy as the neutron to proton ratio in a nuclear system.
We propose the U-spin symmetry energy $E_\mathrm{U}(n_\mathrm{b})$ for hyperonic matter to characterize the variation of binding energy with the inclusion of hyperons. In particular,
we include $\Lambda$ hyperons in dense matter and neglect contributions from other hyperons, then constrain $E_\mathrm{U}(n_\mathrm{b})$ using nuclear physics and astrophysical observations through a Bayesian inference approach.
Our results show that $E_\mathrm{U}(n_\mathrm{b})$ is much smaller than 9$E_\mathrm{sym}(n_\mathrm{b})$, indicating a substantially weaker dependence of the adopted binding energy functional on the U-spin asymmetry than on the corresponding I-spin asymmetry. Consequently, the $\Lambda$ hyperon potential increases significantly with density and becomes repulsive at higher densities, leading to more than 50\% probability for the emergence of $\Lambda$ hyperons in the posterior, which are likely to vanish again at densities $n_\mathrm{b} \gtrsim 5\,n_0$. In scenarios where $\Lambda$ hyperons do emerge, the onset density $n_{\mathrm{b}}^\Lambda$ typically falls in the range $2\,n_0$--$5\,n_0$, corresponding to central densities of neutron stars more massive than $1.0\,\rm{M_\odot}$.
\end{abstract}

\section{\label{introduction}Introduction}

The SU(3) flavor symmetry of light quarks represents a fundamental framework in hadronic physics, which can be decomposed into SU(2)$_I\times$U(1)$_Y$ or SU(2)$_U\times$U(1)$_Q$ subgroups through combinations of ($u$,$d$) I-spin doublets or ($d$,$s$) U-spin doublets~\cite{Greiner1994}. Together with ($u$,$s$) V-spin doublets, these subgroup structures have been extensively utilized to systematically categorize hadrons and characterize their decay properties~\cite{Darewych:1983yw,Julia-Diaz:2007mae,Jia:2019zxi,Zhang:2025jnw}.


The mass differences for hadrons in an I-spin multiplet are typically on the order of 1 MeV, i.e., isospin symmetry is weakly broken. In the context of nucleon interactions, the I-spin symmetry has been thoroughly investigated, where the binding energy per nucleon $E$ for nuclear matter can be divided into symmetric and asymmetric parts, i.e.,
\begin{equation}
  E = E_0(n_\mathrm{b}) + E_\mathrm{sym}(n_\mathrm{b})\delta^2. \label{Eq:EN}
\end{equation}
The first term $E_0(n_\mathrm{b})$ represents the binding energy of symmetric nuclear matter (SNM) and the second one $E_\mathrm{sym}(n_\mathrm{b})$ the nuclear symmetry energy, where $n_\mathrm{b}=n_n + n_p$ is the baryon number density and $\delta= (n_n - n_p)/ (n_n + n_p)$ the asymmetry parameter with $n_{n}$ and $n_{p}$ the neutron and proton number densities. The symmetry energy $E_\mathrm{sym}(n_\mathrm{b})$ quantitatively describes the variation of binding energy with the neutron-to-proton ratio in nuclear systems, which is well constrained at $n_\mathrm{on}=0.1\ \mathrm{fm}^{-3}$ and nuclear saturation density $n_0=0.16\ \mathrm{fm}^{-3}$ with $E_\mathrm{sym}(n_\mathrm{on}) =25.5 \pm 1.0$ MeV~\citep{Centelles:2008vu,Brown:2013mga} and $E_\mathrm{sym}(n_{0}) = 31.7 \pm 3.2$ MeV~\citep{Li:2013ola,Oertel:2016bki}.

Meanwhile, U-spin symmetry in hadrons is explicitly broken by the mass difference in $d$ and $s$ quarks, while the charge symmetry is respected. In such cases, the decay properties involving strangeness can be understood assuming U-spin symmetry~\citep{Darewych:1983yw,Julia-Diaz:2007mae,Jia:2019zxi,Zhang:2025jnw}. Building upon this foundation, we propose in this work a U-spin symmetry energy $E_\mathrm{U}(n_\mathrm{b})$ for hyperonic matter, serving as an analogous quantity that characterizes the binding energy variation upon the inclusion of hyperons.

Being the lightest hyperon, $\Lambda$ hyperons have been investigated extensively, where a large number of $\Lambda$ hypernuclei are produced~\citep{Hashimoto:2006aw,KEKE176:2009jzw, E373KEK-PS:2013dfg,Gal:2016boi}. By reproducing the binding energies of $\Lambda$ hypernuclei, the $N$-$\Lambda$ and $\Lambda$-$\Lambda$ interactions can be fixed based on various nuclear structure models~\citep{Gal:1971gb,Dalitz:1978hw,Millener:2008zz,Millener:2013mwa,Motoba:1983kbv,Hiyama:2009ki,Bando:1990yi,Isaka:2013sfa,Hu:2013tma,Brockmann:1977es,Boguta:1980wi,Mares:1993xb,Sugahara:1994pn,Song:2010qhk,Tanimura:2011ji,Liu:2018img,Rong:2021bim,Rong:2025wnv,Zhou:2007zze,Tsushima:1997cu,Guichon:2008zz}. It was found that a $\Lambda$ potential well depth $V_\Lambda(n_0) = - 30$~MeV in SNM is required to accommodate the single $\Lambda$ binding energies~\citep{Sun:2017fnf,Rong:2021bim}. The study of hyperons in dense matter has attracted significant attention due to their potential role in neutron star interiors, which typically emerge at $\sim$2-4$\,n_0$~\citep{Sun:2017fnf,Sun:2018tmw}. In particular, with the emergence of hyperons, the equation of state (EOS) becomes so soft that consequently the maximum mass of hyperonic stars can not reach the observational two-solar-mass limit~\citep{Demorest:2010bx,Antoniadis:2013pzd,Fonseca:2016tux}, i.e., the hyperon puzzle~\citep{Vidana:2015faa}. Extensive efforts were made to resolve the hyperon puzzle~\citep{Weissenborn:2011ut,Bednarek:2011gd,Oertel:2014qza,Maslov:2015msa,Maslov:2015wba,Takatsuka:2002adv,Vidana:2010ip,Yamamoto:2013ada,Lonardoni:2014bwa,Togashi:2016fky,Weissenborn:2011qu,Klahn:2013kga,Zhao:2015rta,Kojo:2014rca,Masuda:2015kha,Li:2015ida,Whittenbury:2015ziz,Fukushima:2015bda,Sun:2017fnf}. Recent advances in chiral effective field theory ($\chi$EFT) have provided new insights into hyperon potentials~\citep{Lonardoni:2014bwa,Gerstung:2020ktv}, which revealed that repulsive three-body forces between $\Lambda$ hyperons and nucleons play a crucial role in resolving the hyperon puzzle in neutron stars. Their analyses indicate that strongly repulsive $\Lambda$ potentials at high densities effectively suppress hyperon appearance in massive neutron stars.

In this work, we focus on $\Lambda$ hyperons as the lightest hyperonic constituents in dense matter and neglect contributions from other hyperons. The binding energy for SNM $E_0(n_\mathrm{b})$, the nuclear symmetry energy $E_\mathrm{sym}(n_\mathrm{b})$, as well as the U-spin symmetry energy $E_\mathrm{U}(n_\mathrm{b})$ are determined through state-of-the-art constraints from nuclear physics and astrophysical observations employing a Bayesian inference approach~\citep{Xie:2019sqb,Xie:2020tdo,Xie:2023llh,Li:2024imk,Malik:2022zol,Malik:2022ilb,Huang:2023grj,Li:2024pts}. Our analysis reveals that $E_\mathrm{U}(n_\mathrm{b})$ exhibits significantly smaller magnitude compared to its I-spin counterpart  9$E_\mathrm{sym}(n_\mathrm{b})$, indicating that the adopted phenomenological binding energy functional exhibits a substantially weaker dependence on the U-spin asymmetry than on the corresponding I-spin asymmetry.

The paper is organized as follows. In Sec.~\ref{sec:the}, we detail the theoretical framework and computational methods. Section~\ref{sec:res} presents our main results and discusses their implications for neutron star physics. Finally, we summarize our conclusions in Sec.~\ref{sec:con}.

\section{\label{sec:the} Theoretical framework}
To describe hyperonic matter, a general form for the energy density of hyperonic matter can be given as
\begin{eqnarray}
    \varepsilon_\mathrm{b} &=& n_\mathrm{b} E(n_\mathrm{b}, \delta, \delta_\mathrm{U})  +    n_N (n_\mathrm{b}, \delta, \delta_\mathrm{U}) M_{N} \nonumber \\
   {}&&+ n_{\Lambda} (n_\mathrm{b}, \delta, \delta_\mathrm{U}) M_\Lambda, \label{Eq:E_expand1}
\end{eqnarray}
where $M_{N}=939$ MeV and $M_{\Lambda}=1115.6$ MeV are the rest masses of nucleons and $\Lambda$ hyperons, and $E(n_\mathrm{b}, \delta, \delta_\mathrm{U})$ the binding energy per baryon. Nevertheless, Eq.~(\ref{Eq:EN}) is no longer valid with the additional strangeness degrees of freedom. In particular, for a system fulfilling SU(3) symmetry in ($u$,$d$,$s$) space, there are in general eight possible charges. However, since the Cartan subalgebra of SU(3) is two dimensional, it is always possible to transform to a gauge where all are zero except for any two charges represented by two commutative traceless Hermite matrices. We are thus free to choose any two different charges to describe the system, where the third component of I-spin $I_3$ and hypercharge $Y$ are typically employed. In this work, since U-spin respects charge symmetry while V-spin does not, we choose $I_3$ (I-spin) and $U_3$ (U-spin) which are associated with the diagonal generators $\frac{1}{2}$diag$(-1,1,0)$ and $\frac{1}{2}$diag$(0,1,-1)$ in ($u$,$d$,$s$) space. I-spin ($E_\mathrm{sym}$) and U-spin ($E_\mathrm{U}$) symmetry energies emerge. We can then rewrite expression~(\ref{Eq:EN}) for the binding energy per baryon of hyperonic matter as following,
\begin{equation}
  E = E_0(n_\mathrm{b}) + E_\mathrm{sym}(n_\mathrm{b})\delta^2 + E_\mathrm{U}(n_\mathrm{b})(\delta_\mathrm{U}^2-1),
  \label{Eq:Et}
\end{equation}
where the I-spin asymmetry and U-spin asymmetry are defined as
\begin{eqnarray}
\delta &=& 3 \frac{n_d-n_u}{n_d+n_u} = \frac{n_n - n_p}{n_n + n_p + 2n_\Lambda/3}, \label{Eq:delta}\\
\delta_\mathrm{U} &=&  \frac{n_d-n_s}{n_d+n_s} = \frac{2 n_n + n_p}{2 n_n + n_p + 2n_\Lambda}. \label{Eq:deltaU}
\end{eqnarray}
This means only the third components of I-spin and U-spin are considered, which is sufficient as only $\Lambda$ hyperon is included. In order to account for other octet baryons ($\Sigma$ and $\Xi$), we need to consider more carefully the symmetry energy terms, e.g., dividing Eq.~(\ref{Eq:Et}) into kinetic and symmetry terms~\cite{Dover1993_NPA560-559}.
Note that a factor 3 is included in the expression of I-spin asymmetry to be consistent with what was previously defined in Eq.~(\ref{Eq:EN}), where Eq.~(\ref{Eq:Et}) is restored into Eq.~(\ref{Eq:EN}) if we take $n_\Lambda=0$. Here the symmetric ($\Lambda$-only, $\delta=\delta_{\rm U}=0$) matter term ($E_0 - E_\mathrm{U}$) is considered as flavor independent, $E_\mathrm{sym}$ accounts for the energy difference between symmetric nuclear matter and pure neutron matter, and $E_\mathrm{U}$ represents the energy difference between pure $\Lambda$ matter and symmetric nuclear matter, which can be converted into V-spin symmetry energy $E_\mathrm{V}$ by taking $\delta_\mathrm{U} = (\delta+3 \delta_\mathrm{V})/(3+\delta \delta_\mathrm{V})$ in Eq.~(\ref{Eq:Et}). The quadratic form of binding energy
in the U-spin asymmetry $\delta_\mathrm{U}$ is in direct analogy with the empirically established parabolic law for isospin asymmetry $\delta$, which is expected at ultrahigh densities approaching asymptotic freedom. At smaller densities with nonperturbative contributions, however, the mass difference between $\Lambda$ hyperons and nucleons explicitly breaks the SU(3) symmetry and is taken into account with the mass terms in Eq.~(\ref{Eq:E_expand1}), while Eq.~(\ref{Eq:Et}) only accounts for the kinetic and interaction energy.


In principle, we could consider the mass difference as part of $E_\mathrm{U}$ and adopt the current masses of quarks, while the symmetry breaking terms ($\delta_\mathrm{U}$, $\delta_\mathrm{U}^3$, ...) should also be included. Nevertheless, as will be illustrated in Sec.~\ref{sec:res}, the constrained $\delta_{\rm U}$ for neutron star matter is generally close to 1 ($\gtrsim0.9$). If we take $x_{\rm U}=\delta_{\rm U}-1$, the U-spin symmetry energy term in Eq.~(\ref{Eq:Et}) is converted into $E_\mathrm{U}(n_\mathrm{b})(2x_{\rm U} + x_{\rm U}^2)$, i.e., the leading order contribution is accounted for at $\delta_{\rm U}\sim 1$. Including linear, cubic, and higher-order terms in $\delta_{\rm U}$ could in principle alter slightly the onset density and abundance for  $\Lambda$ hyperons. However, as we adopt the piecewise scheme for $E_\mathrm{U}(n_\mathrm{b})$ with a sufficiently large number of segments, the model is flexible enough to describe hyperonic matter in neutron stars. A more detailed investigation including other terms in $\delta_{\rm U}$ should be addressed in our future work.

As was done in previous investigations~\citep{Margueron:2017eqc,Cai:2021ucx,Li:2024imk}, we expand $E_0(n_\mathrm{b})$,  $E_\mathrm{sym}(n_\mathrm{b})$, and $E_\mathrm{U}(n_\mathrm{b})$ in Taylor series around the nuclear saturation density $n_0$, i.e.,
\begin{eqnarray}
E_0 &=& E_0(n_0) + \frac{K_0}{2} x^2 +\frac{J_0}{6} x^3 + \frac{Z_0}{24} x^4, \label{Eq:E0}\\
E_\mathrm{sym} &=&  E_\mathrm{sym}(n_0) + L_{\mathrm{sym}}x+\frac{K_{\mathrm{sym}}}{2}x^2+\frac{J_{\mathrm{sym}}}{6}x^3, \label{Eq:Esym} \\
E_\mathrm{U} &=&  E_\mathrm{U}(n_0) + L_{\mathrm{U}}x+\frac{K_{\mathrm{U}}}{2}x^2 +\frac{J_{\mathrm{U}}}{6}x^3, \label{Eq:EU}
\end{eqnarray}
with $x\equiv \frac{n_\mathrm{b}-n_0}{3\,n_0}$. To improve the model's flexibility,  at densities $n_\mathrm{b} \geq 1.5\,n_0$, we further employ a piecewise Taylor-expanded form for $E_0(n_\mathrm{b})$,  $E_\mathrm{sym}(n_\mathrm{b})$, and $E_\mathrm{U}(n_\mathrm{b})$, where the relevant density range is divided into several segments  and $n_0$ is replaced by $n_i$ in Eqs.~(\ref{Eq:E0}-\ref{Eq:EU}) for the $i$th segment ($n_{i} < n_\mathrm{b} \leq n_{i+1}$). The density breakpoints $n_i$ are treated as random parameters within the range $n_{i-1} < n_i < 3\, n_{i-1}$, whose values are determined by Bayesian inference as illustrated in ~\ref{sec:bayes}. To ensure continuity of the energy, pressure, chemical potential, and sound speed, the zeroth-, first-, and second-order coefficients at the breakpoints $n_i$ are continuous, leaving only the third-order coefficients ($J_{0,i}$, $J_{\mathrm{sym},i}$, $J_{\mathrm{U},i}$) as free parameters, which will be constrained
from advanced nuclear physics input and most recent astrophysical observations. The convergence of our approach is examined by dividing the density range into 1-7 segments employing aforementioned constraints
while the EOS presented by~\cite{Huth:2021bsp} can be reproduced as well in the absence of hyperons, i.e., the model dependencies are minimized.

For a given set of coefficients, the energy density of hyperonic matter can be rewritten as:
\begin{eqnarray}
\varepsilon_\mathrm{b} &=& n_\mathrm{b} E(n_\mathrm{b}, \delta, \delta_\mathrm{U})  - \frac{2 n_\mathrm{b} \left( \delta - \delta \delta_\mathrm{U} -6 \delta_\mathrm{U} \right) }{\delta - \delta \delta_\mathrm{U} +3 \delta_\mathrm{U}+9}   M_{N} \nonumber \\
   {}&&+ \frac {3 n_\mathrm{b} \left(\delta - \delta \delta_\mathrm{U} -3 \delta_\mathrm{U}+3 \right) }{\delta - \delta \delta_\mathrm{U} +3 \delta_\mathrm{U}+9} M_\Lambda, \label{Eq:E_expand}
\end{eqnarray}
where the binding energy $E(n_\mathrm{b}, \delta, \delta_\mathrm{U})$ is fixed by Eq.~(\ref{Eq:Et}).

To fix the EOSs of neutron star matter, the contributions of leptons should be considered, where the total energy density $\varepsilon$ of $np\Lambda e\mu$ matter reads
\begin{equation}
  \varepsilon =\varepsilon_\mathrm{b}(n_\mathrm{b}, \delta, \delta_\mathrm{U}) + \sum_{l=e,\mu}\varepsilon_{l}(n_l). \label{Eq:Ett}
\end{equation}
Here $\varepsilon_l(n_l)={m_l^{4}} f\left({\sqrt[3]{3\pi^2 n_l}}/{m_l}\right)/{8\pi ^2}$ is the energy density of leptons as a function of the number density of leptons $n_l$,
where $f(y)\equiv \left[y\sqrt{1+y^2}\left(1+2y^2\right)-\mathrm{arcsh}(y)\right]$, $m_e=0.511$ MeV and $m_\mu=105.66$ MeV refer to the electron and muon masses. The chemical potential of particle type $i$ can be estimated with
\begin{equation}
\mu_i=\left.\frac{\partial \varepsilon}{\partial n_i}\right|_{n_{j\neq i}}.
\end{equation}
The pressure is then obtained with
\begin{equation}
  P = \sum_i n_i \mu_{i} - \varepsilon. \label{Eq:Pt}
\end{equation}
We can also fix the potential depth of $\Lambda$ hyperons $V_\Lambda$ in nuclear medium ($n_\Lambda=0$) by
\begin{eqnarray}
V_\Lambda  &=& \mu_\Lambda  - M_\Lambda \label{Eq:VL} \\
    &=&  \frac{\mbox{d}E_0}{\mbox{d}n_\mathrm{b}} n_\mathrm{b} +E_0 +
         \frac{\mbox{d}E_\mathrm{sym}}{\mbox{d}n_\mathrm{b}} n_\mathrm{b}\delta^2 \nonumber \\
    &&{} - \frac{1}{3} E_\mathrm{sym}\delta^2
         -\frac{ 8E_\mathrm{U}}{\delta+3}. \nonumber
\end{eqnarray}
where \(\delta\) is defined in Eq.~(\ref{Eq:delta}).

Through the $\beta$-equilibrium and charge neutrality conditions
\begin{eqnarray}
\mu _{e}&=&\mu_{\mu }=\mu_{n}-\mu_{p}=\mu_\Lambda-\mu_{p}, \label{Eq:chem_equiv} \\
n_{p}&=&n_{e}+n_\mu = \frac{2n_\mathrm{b} \left(3 \delta_\mathrm{U} -\delta \delta_\mathrm{U} - 2 \delta \right) }{\delta-\delta \delta_\mathrm{U}  + 3 \delta_\mathrm{U}+9}, \label{Eq:charge}
\end{eqnarray}
we can obtain the isospin asymmetry $\delta(n_\mathrm{b})$, U-spin asymmetry $\delta_\mathrm{U}(n_\mathrm{b})$, and relative particle fractions ($n_i/n_\mathrm{b}$ with $i=p,n,\Lambda,e,\mu$) of neutron star matter at fixed baryon number density $n_\mathrm{b}$, while the speed of sound squared is determined by
\begin{equation}
 c_s^2 = \frac{\mbox{d}P}{\mbox{d}\varepsilon}.
\end{equation}
The compact star structure is then fixed by solving the Tolman-Oppenheimer-Volkoff (TOV) equations
\begin{eqnarray}
\frac{\mbox{d}P}{\mbox{d}r} & = & -\frac{G M {\varepsilon}}{r^2} \frac{(1 + P/\varepsilon)(1 + 4\pi r^3 P/M)}{1 - 2GM/r}, \label{eq:TOV} \\
\frac{\mbox{d}M}{\mbox{d}r} & = & 4\pi {\varepsilon} r^2, \label{eq:m_star}
\end{eqnarray}
with the gravitational constant $G$ = $6.707 \,\times 10^{-45}$ \text{MeV}$^{-2}$. Note that at $n_\mathrm{b}< 0.08\ \mathrm{fm}^{-3}$ we employ the crust EOS predicted by the relativistic density functional TW99 with the slope of symmetry energy $L_\mathrm{sym}(n_0)$ close to the peak value obtained here~\citep{Xia:2022dvw}.

\section{\label{sec:res}Results and discussions}

\begin{figure*}[t!]
\centering
\includegraphics[width=0.9\linewidth]{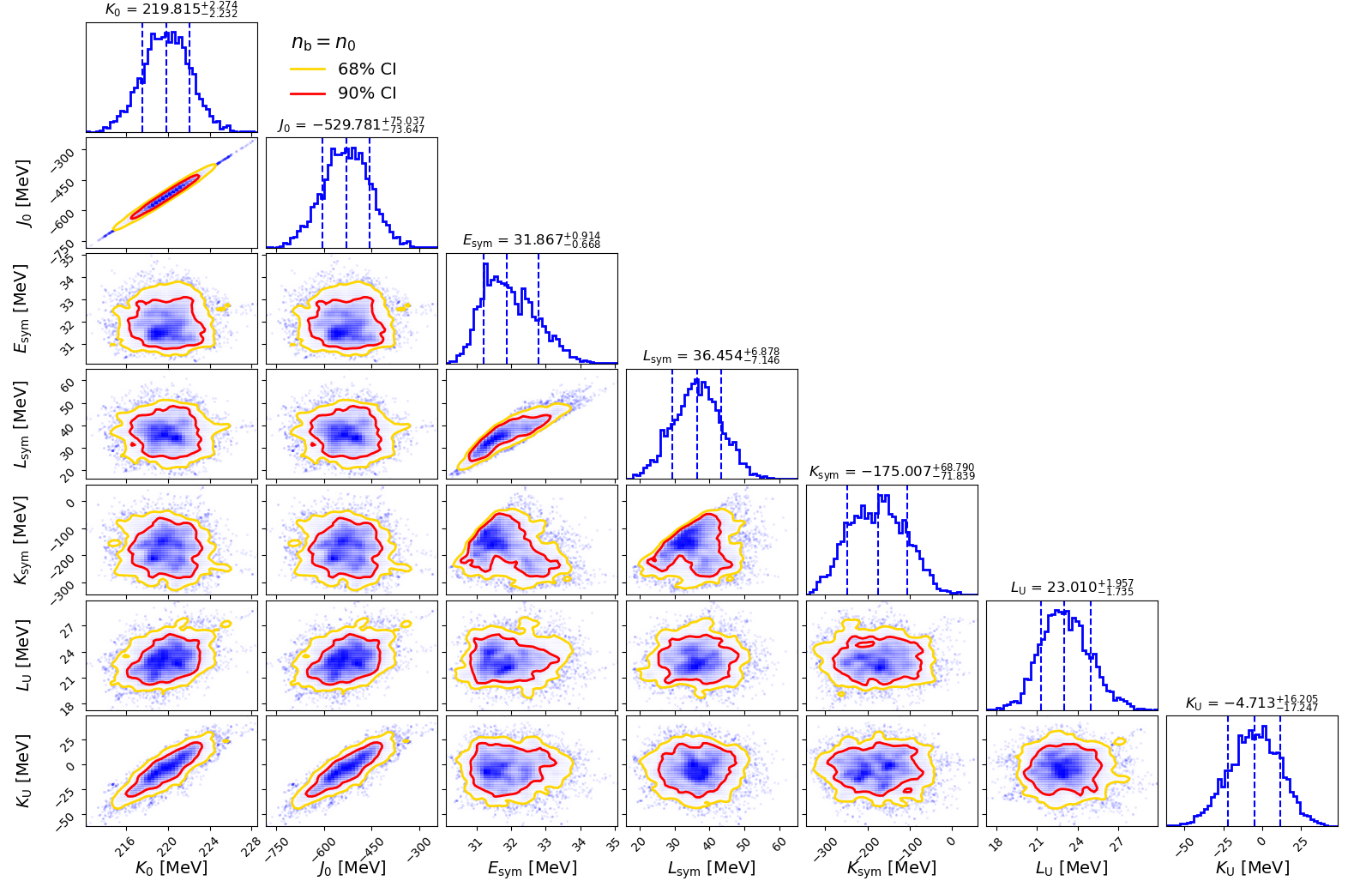}
\caption{\label{Fig:para1} PDFs of the saturation properties as well as their correlations inferred from the Bayesian analysis of both nuclear physics constraints and astrophysical constraints listed in Table~\ref{Tab:HIC}, Table~\ref{Tab:sym}, Table~\ref{Tab:U-lambda} and Table~\ref{Tab:data_astro}, employing prior distribution as indicated in Table~\ref{Tab:prior} for hyperonic EOSs in this work. The red (yellow) curves indicate the 68\% (90\%) credible regions.}
\end{figure*}

\begin{figure*}[t!]
  \centering
  \includegraphics[width=0.8\linewidth]{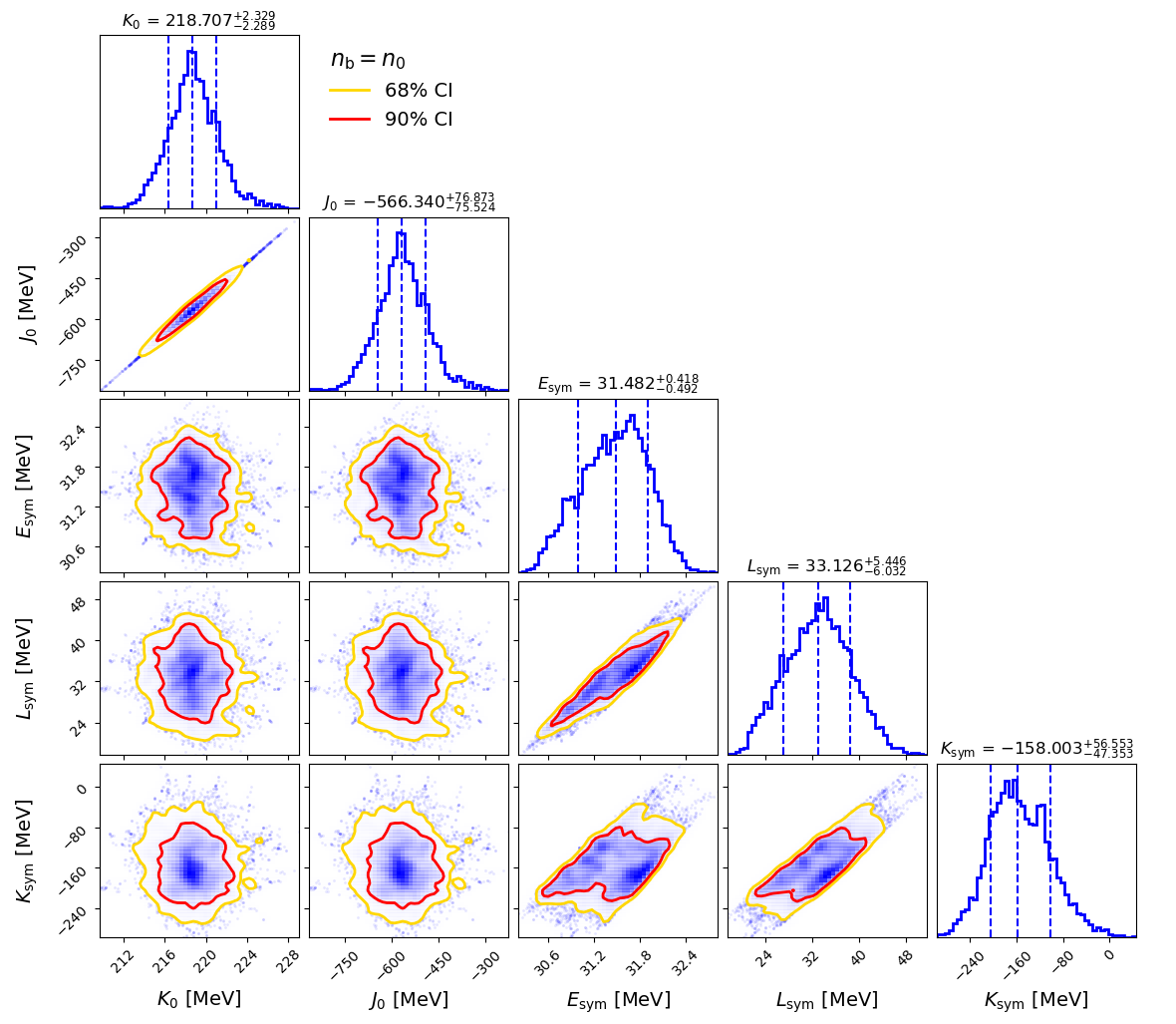}
\caption{\label{Fig:para2} PDFs of the saturation properties and their correlations, inferred from the Bayesian analysis of nuclear physics constraints and astrophysical constraints listed in Table~\ref{Tab:HIC}, Table~\ref{Tab:sym}, and Table~\ref{Tab:data_astro}, without considering the U-spin symmetry energy (i.e., hyperons are excluded). The prior distributions of the relevant parameters are consistent with those in Fig.~\ref{Fig:para1}. The red (yellow) curves indicate the 68\% (90\%) credible regions.}
\end{figure*}

Based on various nuclear and astrophysical constraints, we can then constrain the model parameters and EOSs of hyperonic matter using a Bayesian inference approach, where the details are illustrated in the appendix. In Fig.~\ref{Fig:para1} we present the obtained posterior probability distribution functions (PDFs) on the saturation properties of hyperonic matter and their correlations, i.e., the incompressibility parameter $K_0$, skewness parameter $J_0$, symmetry energy $E_\mathrm{sym}(n_0)$, slope and curvature of symmetry energy $L_\mathrm{sym}$ and $K_\mathrm{sym}$, slope and curvature of U-spin symmetry energy $L_\mathrm{U}$ and $K_{\mathrm{U}}$, where the 68\% and 90\% credible regions are indicated as well. As illustrated in Sec.\ref{sec:the}, the binding energy $E_{0}(n_0) = -16$ MeV, $E_0 (n_{\mathrm{on}}) = -14.1$ MeV, the symmetry energy $E_{\mathrm{sym}} (n_{\mathrm{on}}) = 25.5$ MeV and U-spin symmetry energy $E_\mathrm{U}(n_0) =  5.25$ MeV (corresponding to $V_\Lambda(n_0) = -30$ MeV in SNM) take constant values, while $J_0$ and  $K_\mathrm{sym}$ are connected to the parameters $K_0$, $E_\mathrm{sym} (n_0)$ and $L_{\mathrm{sym}}$ via $J_0 = 33K_0-7783.68$ MeV and $K_{\mathrm{sym}} = 22 L_{\text{sym}} - 194 E_{\text{sym}}(n_0) + 5222$ MeV.

As indicated in Fig.~\ref{Fig:para1}, it is evident that the obtained PDFs of $K_0$, $J_0$, $E_\mathrm{sym}(n_0)$, $L_\mathrm{sym}$, $K_\mathrm{sym}$, $L_{\mathrm{U}}$ and $K_{\mathrm{U}}$ have a single peak structure, suggesting that they are well constrained according to both nuclear and astrophysical data. Note that a Gaussian prior is assumed for the incompressibility parameter with $K_0=240\pm20$ MeV~\citep{Shlomo:2006ole}. The fact that its peak value lies near the lower $1\sigma$ limit of the prior distribution suggests that both nuclear and astrophysical constraints favor a smaller $K_0$. The obtained constraints on $K_0$ and $J_0$ are generally consistent with previous estimations~\citep{Shlomo:2006ole,Xie:2020kta}, while that of nuclear symmetry energy $E_{\mathrm{sym}}(n_0)$, its slope $L_{\mathrm{sym}}$ and curvature $K_{\mathrm{sym}}$ are also consistent with those derived from the isospin diffusion data in HICs~\citep{Tsang:2008fd}, neutron skin thickness, and neutron star structures~\citep{Zhang:2020azr}, despite that the constrained value of $L_{\mathrm{sym}}$ is relatively small. Strong positive correlations are observed among $K_0$ and $J_0$, which are attributed to our model assumptions, i.e., $J_0 = 33K_0-7783.68$ MeV. Additionally, positive correlations for the $E_{\mathrm{sym}}(n_0)$--$L_{\mathrm{sym}}$ and $K_\mathrm{U}$--$K_0$/$J_0$ pairs are observed as well, which are derived from hyperonic EOS model constraints with the explicit inclusion of U-spin symmetry energy.

At the saturation density, \(E_{\rm U}(n_0)\) is much smaller in magnitude than \(9E_{\rm sym}(n_0)\), indicating that the adopted phenomenological energy functional has a substantially weaker dependence on the U-spin asymmetry than on the corresponding I-spin asymmetry. For the density-dependent behavior of $E_{\mathrm{U}}(n_{\mathrm{b}})$, the slope parameter $L_{\mathrm{U}}$ is well constrained, with a value of $23.010^{+1.957}_{-1.755} \mathrm{MeV}$, and exhibits a pronounced localized peak in the posterior distribution. This arises from the constraints imposed by the $\Lambda$ potential well depths in SNM at different densities according to the SU(3) $\chi$EFT incorporating three-body forces~\citep{Gerstung:2020ktv}, which directly affect the first
3 segments of our hyperonic EOS model. Under the current constraints, where the curvature is also determined as $K_{\mathrm{U}} = -4.713^{+16.205}_{-17.247}$ MeV, the distribution peaks near zero, implying a nearly linear growth of $E_{\mathrm{U}} (n_0)$. As density increases, both the neutron chemical potential $\mu_n$ and $\Lambda$ chemical potential $\mu_\Lambda$ rise, and the competition between them determines the onset of hyperons. In fact, as indicated in Table~\ref{tab:MR_summary}, there is only 55.66\% probability that $\Lambda$ hyperons emerge in neutron stars matter within the density range considered here. For hyperon-bearing neutron stars, their masses should at least exceed $1.0\, \rm{M_{\odot}}$, while the onset density for hyperons $n_{\mathrm{b}}^{\Lambda}$ generally ranges from $2\,n_0$ to $5\,n_0$. Furthermore, the emergence of hyperons triggers the hyperonic direct Urca process, which significantly accelerates the cooling processes of neutron stars~\citep{Prakash:1992zng}.

\begin{figure}[ht!]
  \centering
  \includegraphics[width=\linewidth]{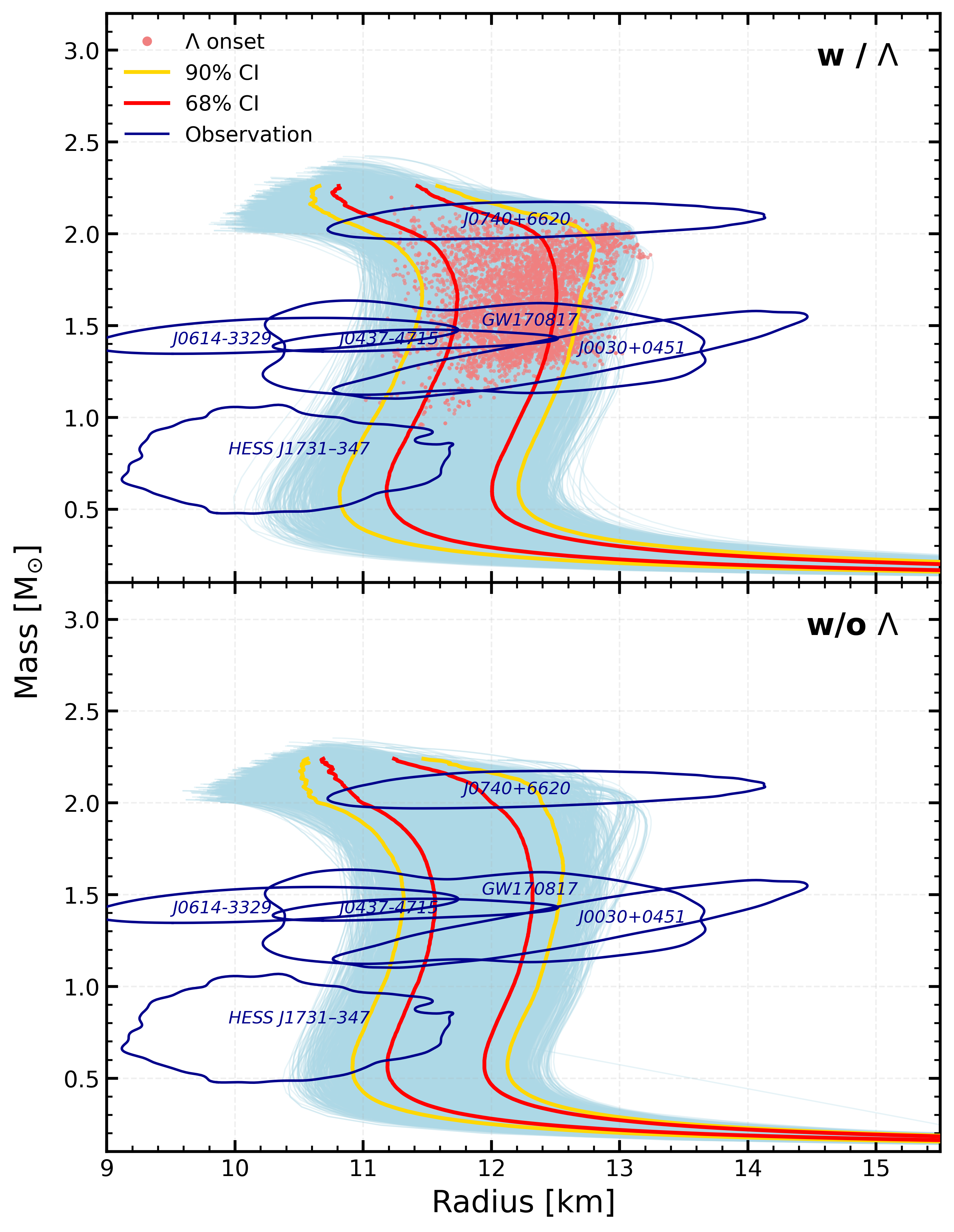}
  \caption{\label{Fig:mr}Mass-radius relations of neutron stars predicted using the parameters shown in Fig.~\ref{Fig:para1} (upper panel) and Fig.~\ref{Fig:para2} (lower panel), adopting the EOS models with and without $\Lambda$ hyperons, respectively. The posterior 68\% (red) and 90\% (yellow) credible regions for neutron star masses and radii from the Bayesian analysis are also presented. Red filled circles denote the critical mass of neutron stars that triggers the onset of $\Lambda$ hyperons in their centers. }
\end{figure}

\begin{table*}
\centering
\renewcommand{\arraystretch}{1.5} %
\footnotesize 
\setlength{\tabcolsep}{3pt} %
\caption{Statistical results of neutron star properties at 5 representative masses $M$ in the upper panel of Fig.~\ref{Fig:mr}. The table lists the 68\% and 90\% credible intervals for the radius $R$, central baryon density $n_{\rm b}^{\rm cent}$, pressure $P^{\rm cent}$, and energy density $\varepsilon^{\rm cent}$. The last three columns show the percentages of samples satisfying the proton fraction constraint ($f_p > 14.8\%$)~\citep{Lattimer:1991ib} of nucleonic direct Urca processes (with muons present in the system), the $\Lambda$ hyperon onset constraint ($\delta_\mathrm{U} < 1$), and either one of the two constraints ($f_p > 14.8\%$ or $\delta_\mathrm{U} < 1$), which correspond to the probability for the onset of nucleon Urca process ($p_\mathrm{NDU}$), hyperon Urca process ($p_\mathrm{HDU}$), and either one of the two types of Urca processes ($p_\mathrm{DU}$).}
\label{tab:MR_summary}
\begin{tabular}{lccccccccccc}
\hline\hline
$M$ ($\mathrm{M}_\odot$) & \multicolumn{2}{c}{$R$ (km)} & \multicolumn{2}{c}{$n_{\text{b}}^{\rm cent}$ (fm$^{-3}$)} & \multicolumn{2}{c}{$P^{\rm cent}$ (MeV/fm$^{3}$)} & \multicolumn{2}{c}{$\varepsilon^{\rm cent}$ (MeV/fm$^{3}$)} &  $p_\mathrm{NDU}$ & $p_\mathrm{HDU}$ & $p_\mathrm{DU}$  \\
 &  68\%  &  90\%  &  68\%  &  90\%  & 68\%  &  90\%  & 68\%  &  90\%  &   & &   \\
\hline
0.5 & $11.70_{-0.48}^{+0.35}$ & $11.70_{-0.86}^{+0.57}$ & $0.28_{-0.03}^{+0.03}$ & $0.28_{-0.04}^{+0.06}$ & $12.55_{-1.55}^{+2.47}$ & $12.55_{-2.17}^{+4.48}$ & $267.9_{-23.3}^{+35.7}$ & $267.9_{-33.5}^{+62.6}$ & 0 & 0 & 0 \\
1.0 & $11.87_{-0.45}^{+0.35}$ & $11.87_{-0.78}^{+0.54}$ & $0.35_{-0.02}^{+0.04}$ & $0.35_{-0.04}^{+0.07}$ & $33.13_{-3.83}^{+6.20}$ & $33.13_{-5.53}^{+11.19}$ & $347.2_{-28.5}^{+40.1}$ & $347.2_{-42.0}^{+71.5}$ & 0 & $0.9 \%$ & $0.9 \%$ \\
1.4 & $12.12_{-0.44}^{+0.33}$ & $12.12_{-0.76}^{+0.51}$ & $0.43_{-0.05}^{+0.04}$ & $0.43_{-0.08}^{+0.08}$ & $61.23_{-8.19}^{+11.32}$ & $61.23_{-12.14}^{+20.21}$ & $427.5_{-46.7}^{+48.7}$ & $427.5_{-78.7}^{+88.1}$ & $17.84\%$ & $11.55\%$ & $26.28\%$ \\
1.8 & $12.09_{-0.41}^{+0.40}$ & $12.09_{-0.67}^{+0.68}$ & $0.54_{-0.08}^{+0.07}$ & $0.54_{-0.14}^{+0.12}$ & $122.0_{-26.5}^{+28.0}$ & $122.0_{-41.8}^{+53.1}$ & $563.9_{-93.4}^{+88.2}$ & $563.9_{-156.0}^{+145.9}$ & $81.65\%$ & $42.92\%$ & $86.61\%$ \\
2.0 & $11.84_{-0.44}^{+0.50}$ & $11.84_{-0.76}^{+0.90}$ & $0.67_{-0.10}^{+0.10}$ & $0.67_{-0.17}^{+0.18}$ & $197.9_{-54.0}^{+68.5}$ & $197.9_{-84.9}^{+126.6}$ & $734.2_{-129.8}^{+136.3}$ & $734.2_{-208.9}^{+251.8}$ & $98.43\%$ & $55.66\%$ & $99.17\%$ \\
\hline\hline
\end{tabular}
\end{table*}

For comparison, we have also carried out calculations in the absence of hyperons, where Eq.~(\ref{Eq:Et}) is restored into Eq.~(\ref{Eq:EN}) as we take $\delta_\mathrm{U}=1$. In particular, we employ the same piecewise scheme and prior distribution ranges as in Fig.~\ref{Fig:para1}, where the obtained PDFs and correlations of the saturation properties of nuclear matter are shown in Fig.~\ref{Fig:para2}. The posterior distributions of $K_0$, $J_0$, $E_{\mathrm{sym}}(n_0)$, $L_{\mathrm{sym}}$, and $K_{\mathrm{sym}}$ all exhibit a single-peak structure. Compared with the posterior distributions in Fig.~\ref{Fig:para1}, almost all constrained coefficients in Fig.~\ref{Fig:para2} are slightly smaller. Notably, the obtained incompressibility parameter $K_0 = 218.707_{-2.289}^{+2.329}$ MeV is consistent with its empirical value~\citep{Shlomo:2006ole,Xie:2020kta}, while the corresponding skewness $J_0$ is smaller than previous estimations~\citep{Xie:2020kta}. Consequently, the nuclear EOSs become softer and show better consistency with the constraint derived from EOS models that extrapolate the speed of sound ($c_s$) at higher densities~\citep{Huth:2021bsp}. Similar to Fig.~\ref{Fig:para1}, we have also observed positive correlations for the $K_0$--$J_0$ and $E_{\mathrm{sym}}(n_0)$--$L_{\mathrm{sym}}$/$K_{\mathrm{sym}}$ pairs, i.e., the emergence of hyperons breaks the $L_{\mathrm{sym}}$--$K_{\mathrm{sym}}$ correlation~\citep{Guan:2025yia}.
\begin{figure*}[t]
    \centering
    \includegraphics[width=\textwidth]{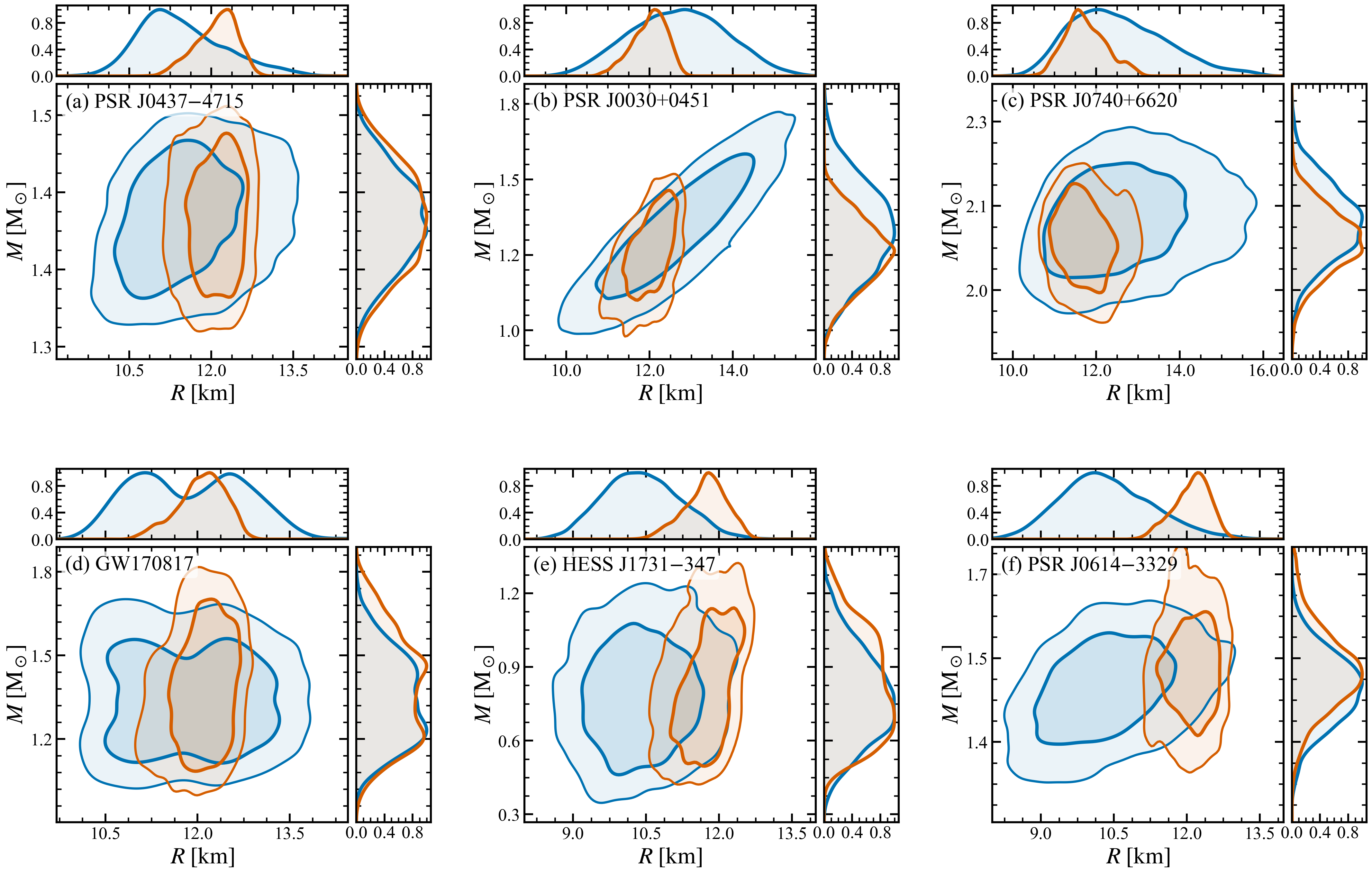}
    \caption{Comparison between the observational constraints and the posterior mass-radius distributions for the six sources listed in Table~\ref{Tab:data_astro}. The blue and orange distributions represent the observational constraints and the posterior distributions obtained from the full joint Bayesian analysis with $\Lambda$. The inner and outer contours enclose the 68\% and 95\% credible regions in the mass-radius plane. For each source, the upper and right panels show the corresponding normalized one-dimensional marginal distributions of radius and mass, respectively.}
    \label{fig:MR_six_sources}
\end{figure*}

By fulfilling the $\beta$-equilibrium and charge neutrality conditions in Eqs.~(\ref{Eq:chem_equiv}) and (\ref{Eq:charge}), the EOSs of neutron star matter are determined through Eqs.~(\ref{Eq:Et}) and (\ref{Eq:Pt}). Subsequently, the mass and radius of a neutron star can be obtained by solving the
TOV equations~(\ref{eq:TOV}) and (\ref{eq:m_star}) at a given central energy density $\varepsilon_c$. Fig.~\ref{Fig:mr} illustrates the mass-radius ($M$-$R$) relations calculated using the parameter sets from Fig.~\ref{Fig:para1} (including $\Lambda$ hyperons) and Fig.~\ref{Fig:para2} (excluding $\Lambda$ hyperons). The posterior 68\% (red) and 90\% (yellow) credible regions of the neutron star mass and radius from the Bayesian analysis are also presented. Regardless of whether $\Lambda$ hyperons are included, the overall trends of the $M$-$R$ relations remain nearly identical. For low-mass ($\sim 1\,\rm{M_\odot}$) neutron stars, both cases predict relatively small radii, which is driven by the small radius measured with large uncertainty for the central compact object (CCO) in the supernova remnant HESS J1731-347~\citep{Doroshenko:2022nwp}. Regarding the predictions for a $1.4 \,\rm{M_\odot}$ neutron star, the mass-radius measurement of PSR J0614-3329~\citep{Mauviard:2025dmd} contributes little to constraining the model space. This is mainly because its small radius measured with large uncertainty exhibits a significant discrepancy with previous results obtained from GW170817\citep{LIGOScientific:2018cki}, PSR J0030+0451~\citep{Riley:2019yda} and PSR J0437-4715~\citep{Choudhury:2024xbk}, which are within the same mass range. In the upper panel of Fig.~\ref{Fig:mr}, the red filled circles indicate the critical mass where $\Lambda$ hyperons begin to emerge at the center of a neutron star. As illustrated also in Table~\ref{tab:MR_summary}, their masses are typically greater than $1.0\, {\rm M}_\odot$. In our posterior EOS statistics, the probability that hyperons appear in the EOSs within the causality-satisfying density range is 55.66\%. For some posterior EOSs, a direct extrapolation of piecewise Taylor-expansion energy functional to densities beyond those relevant to a \(1.97\, \rm M_\odot\) neutron star may become acausal at sufficiently high densities due to the lack of constraints on the ultra-high-density EOS. To avoid using such acausal branches, once an EOS reaches the causal limit \(c_s^2=\text{d}P/\text{d}\varepsilon=1\), its higher-density part is replaced by a constant-sound-speed causal continuation, 
\begin{equation}
    P(\varepsilon)=P_t+(\varepsilon-\varepsilon_t),
    \label{eq:constant cs}
\end{equation}
where \((\varepsilon_t,P_t)\) denotes the matching point ($c_s^2=1$). The complete stable neutron star sequence is then reconstructed up to its maximum mass configuration. This treatment preserves the original posterior EOS below the causal limit matching point and replaces only its acausal high-density extrapolation. With this causal completion, the inferred maximum mass is \(M_{\max}=2.146^{+0.071}_{-0.070}\,\rm M_\odot\) at the 68\% credible regions.

To further examine the role of the individual astrophysical measurements, Fig.~\ref{fig:MR_six_sources} compares the observational M--R probability distributions of the six sources in Table~\ref{Tab:data_astro} with the corresponding posterior distributions obtained from the full joint Bayesian analysis posterior with $\Lambda$. The comparison shows that the inferred M--R relation reflects the combined effect of the different measurements rather than being dominated by any single source. HESS J1731--347 provides the most direct observational preference for relatively small radii in the low-mass region, although the corresponding joint posterior is shifted toward somewhat larger radii. Similarly, the posterior associated with PSR J0614--3329 is shifted toward larger radii, closer to the regions favored by PSR J0030+0451, PSR J0437--4715, and GW170817, indicating its relatively limited leverage on the combined posterior.

\begin{figure*}[t!]
  \centering
  \includegraphics[width=0.451\linewidth]{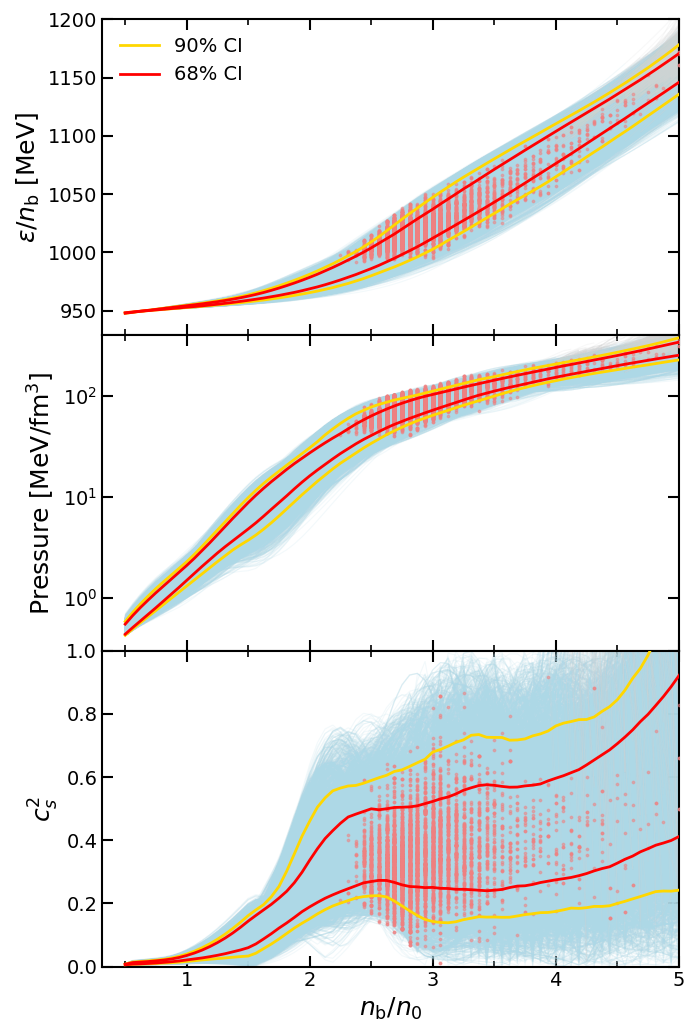} \ \
  \includegraphics[width=0.45\linewidth]{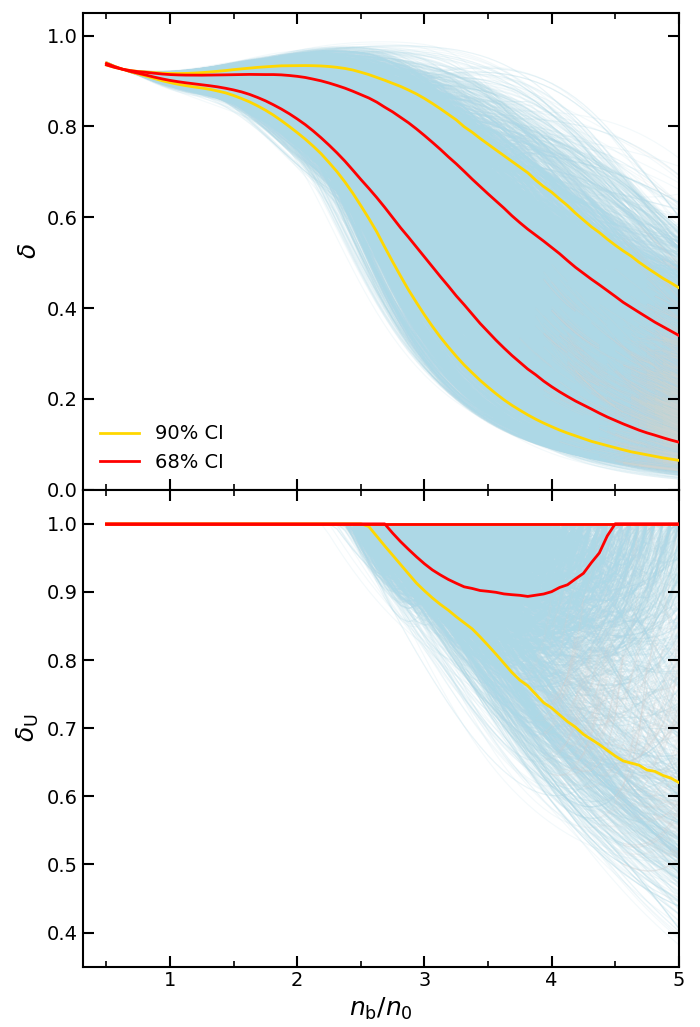}
  \caption{\label{Fig:NSEOS} Left: Energy per baryon $\varepsilon/n_{\rm{b}}$, pressure $P$, and speed of sound squared $c^2_s$ of the neutron star matter; Right: The corresponding I-spin and U-spin asymmetry parameters $\delta$ and $\delta_{\mathrm{U}}$, where the direct Urca processes are triggered once $\delta < 0.704$ given $\delta_{\mathrm{U}}=1$ or at $\delta_{\mathrm{U}}<1$. The red dots represent the onset of $\Lambda$ hyperon. The results presented here are calculated using the parameters of the hyperonic EOS models from Fig.~\ref{Fig:para1}.}
\end{figure*}

Table~\ref{tab:MR_summary} presents the statistical results for the radii and corresponding physical parameters of the EOSs at
5 representative masses, based on the predictions for $\Lambda$-hyperon admixed neutron stars shown in the upper panel of Fig.~\ref{Fig:mr}. From a macroscopic structural perspective, as the mass $M$ increases, the radius $R$ exhibits a non-monotonic behavior -- it first increases and then decreases. On a microscopic level, the central baryon density $n_{\rm b}^{\rm cent}$, pressure $P^{\rm cent}$, and energy density $\varepsilon^{\rm cent}$ all show significant non-linear growth as mass increases. For neutron stars with $M \le 1.0 \,{\rm M}_\odot$, the criterion for nucleonic direct Urca process (including muons) with the proton fraction $f_p  > 14.8\%$~\citep{Lattimer:1991ib}
is not satisfied, and there is only a $0.9\%$ probability for hyperon emergence, i.e., in the absence of hyperonic direct Urca process. Therefore, these low-mass stars cool almost exclusively through the modified Urca process. For a canonical \(1.4\,\rm M_\odot\) neutron star, the probability for either nucleonic or hyperonic direct Urca processes is only \(26.28\%\), while in \(73.72\%\) of the posterior samples neither process is activated. In contrast, for a $1.8 \,{\rm M}_\odot$ neutron star, the probability of satisfaction of the proton abundance constraints $f_p > 14.8\%$ rises sharply, and the likelihood of hyperon onset at high densities also increases significantly, implying a much larger probability for the activation of direct Urca processes in such massive stars. Note that the direct-Urca cooling rates may also be significantly modified by nucleon superfluidity and superconductivity, whose quantitative treatment would require additional microscopic properties not included in the present energy functional.  Analyzing physical properties through these mass slices provides crucial insights into the internal composition and thermal evolution of neutron stars.

Figure~\ref{Fig:NSEOS} illustrates the posterior distribution for EOSs of neutron star matter derived from Bayesian analysis based on the hyperonic EOS model shown in Fig.~\ref{Fig:para1}, alongside the density dependence of the asymmetry parameters $\delta$ and $\delta_\mathrm{U}$. Due to the lack of sufficient constraints from ultra-high densities, a direct extrapolation of piecewise Taylor-
expansion energy functional may become acausal, with \(c_s^2>1\). Consequently, we only present the microscopic properties for densities below \(5\,n_0\). The constant-sound-speed continuation introduced above specifies only the relation between pressure and energy density and does not provide information on the microscopic composition of matter beyond the causal-limit matching point.  The red and yellow curves in the figure represent the 68\% and 90\% credible intervals respectively. Generally, the energy per baryon $\varepsilon/n_{\rm b}$, pressure $P$, and the speed of sound squared $c^2_s$ of neutron star matter increase monotonically with density, with their associated uncertainties expanding accordingly.
There is a 55.66\% probability for the emergence of $\Lambda$ hyperons in the EOS posterior, and the onset of $\Lambda$ hyperons triggers a sharp drop in the speed of sound.
However, within the density range where the piecewise Taylor-
expansion energy functional remains causal, the 68\% credible interval of $\delta_\mathrm{U}$ in the right panel suggests that these hyperons are likely to vanish again with increasing baryon number density $n_\mathrm{b}$. The isospin asymmetry $\delta$ decreases monotonically with density, initially gradually before steepening at higher densities. In the absence of hyperons, the direct Urca process occurs once $\delta < 0.704$ with the proton fraction $f_p > 14.8\%$.
With the inclusion of hyperons, the U-spin asymmetry $\delta_\mathrm{U}$ also decreases with density. Combined with the data in Table~\ref{tab:MR_summary}, it is evident that hyperons could appear before the proton fraction reaches the 14.8\% threshold. This implies that the hyperonic direct Urca process may be activated earlier than the nucleonic direct Urca process. Specifically, within the 68\% credible interval, hyperons begin to emerge at densities above $2.5\,n_0$ and may disappear when the density exceeds $4.5\,n_0$.

\begin{figure}[t!]
  \centering
  \includegraphics[width=\linewidth]{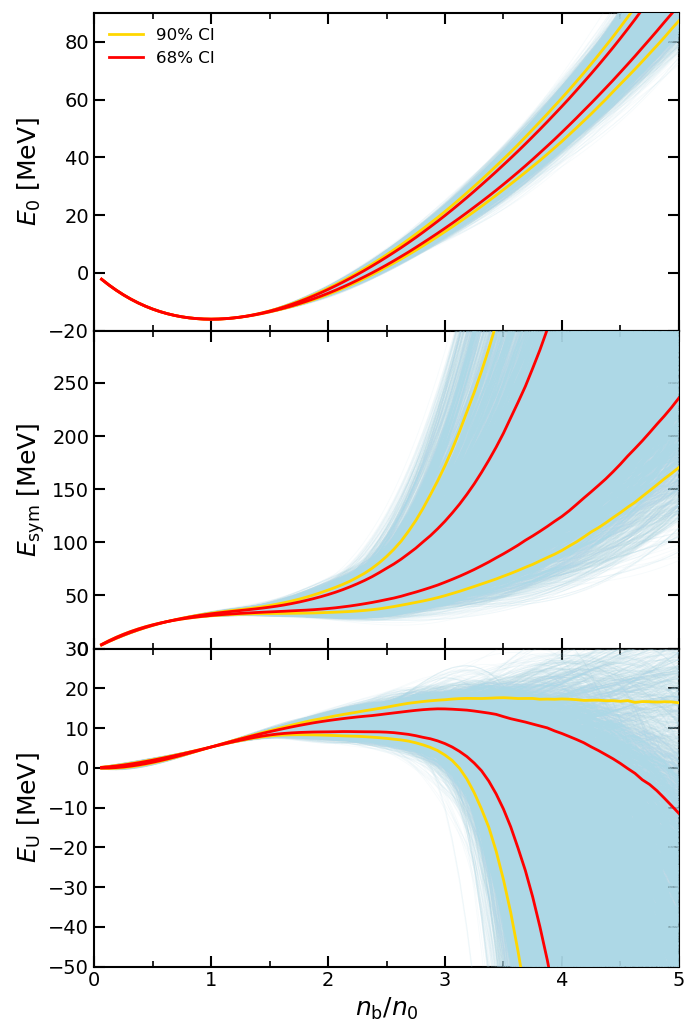}
\caption{\label{Fig:E0-Esym-Eu} Constraints on $E_{0}(n_\mathrm{b})$, $E_\mathrm{sym}(n_\mathrm{b})$, and $E_\mathrm{U}(n_\mathrm{b})$ for Eq.~(\ref{Eq:Et}) employing hyperonic EOS model, where the red (yellow) curves indicate the 68\% (90\%) credible regions.}
\end{figure}

Based on the parameters corresponding to the hyperonic EOS model in Fig.~\ref{Fig:para1}, we present in Fig.~\ref{Fig:E0-Esym-Eu} the binding energy per baryon for SNM $E_{0}(n_{\rm b})$, the nuclear symmetry energy $E_{\rm sym}(n_{\rm b})$, and the U-spin symmetry energy $E_{\rm U}(n_{\rm b})$ as functions of the baryon number density $n_{\rm b}$. These results are obtained using Eqs.~(\ref{Eq:E0}--\ref{Eq:EU}) and the piecewise Taylor-expansion with respect to density. Notably, the U-spin symmetry energy $E_{\rm U}(n_{\rm b})$ is much smaller than the nuclear symmetry energy $9E_{\rm sym}(n_{\rm b})$, and its density dependence differs significantly from that of $E_{\rm sym}(n_{\rm b})$. At $n_{\rm b} \lesssim 3\,n_0$, $E_{\rm U}$ increases with density until reaching a peak with relatively small uncertainties. This behavior favors the appearance of $\Lambda$ hyperons according to the constraint on $\Lambda$ potential well depth in SNM listed in Table~\ref{Tab:U-lambda}. In the high-density region where $n_{\rm b} > 3\,n_0$, $E_{\rm U}$ gradually decreases and tends toward negative values as density increases, implying a high probability that hyperons will disappear at higher densities. This is primarily constrained by astrophysical observations; to satisfy the mass requirement of the massive neutron star PSR~J0740+6620~\citep{Miller:2021qha}, the EOS cannot be too soft to support a $2\,\rm{M_{\odot}}$ configuration. These results indicate that the effective \(\Lambda\) potential becomes progressively less attractive with increasing density and may eventually turn repulsive in the high-density region.

\begin{figure}[t!]
  \centering
  \includegraphics[width=\linewidth]{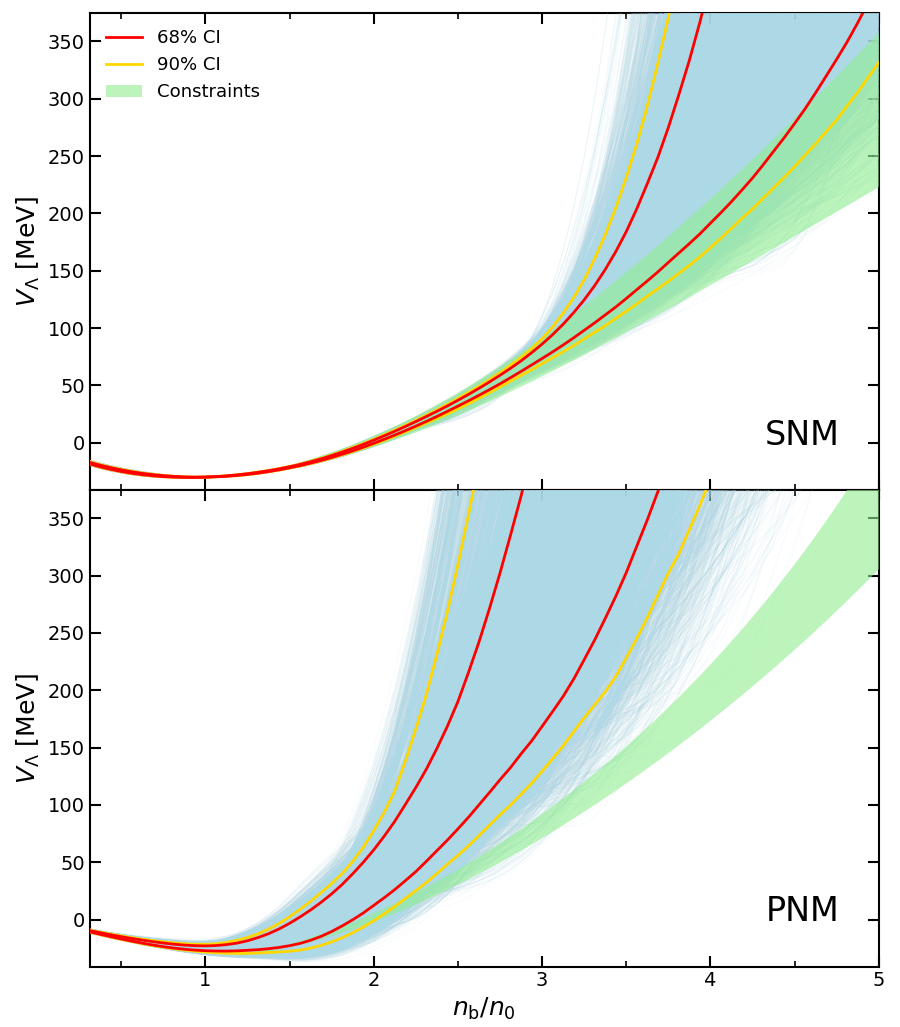}
\caption{\label{Fig:EOS_nucl2} $\Lambda$ potential well depths in SNM and PNM as functions of baryon density $n_\mathrm{b}$, along with their 68\% and 90\% credible intervals. The green shaded region denotes calculations based on the SU(3) $\chi$EFT incorporating three-body forces~\citep{Gerstung:2020ktv}.}
\end{figure}

By taking vanishing densities for leptons and hyperons, i.e., $n_e = n_\mu = n_\Lambda = 0$ (corresponding to $\delta_\mathrm{U} = 1$), we can fix the potential depths of $\Lambda$ hyperons in SNM and PNM by setting $\delta = 0$ and 1. Fig.~\ref{Fig:EOS_nucl2} presents the 68\% and 90\% credible intervals for the $\Lambda$ potential depth $V_\Lambda(n_\mathrm{b})$ predicted by the hyperonic EOS model. Evidently, for $n_{\rm b}\lesssim 2.5\,n_0$, 
although we only utilize the $\Lambda$ potential depth in SNM from Table~\ref{Tab:U-lambda} as a constraint, our calculated $\Lambda$ potential in PNM is also roughly consistent with the results in
Ref.~\cite{Gerstung:2020ktv}. At high densities, the predictions of the SU(3) $\chi$EFT may become unreliable and were not employed in our Bayesian analysis. Consequently, the obtained values on $V_\Lambda(n_\mathrm{b})$ tend to be larger than $\chi$EFT predictions, while the uncertainties increase significantly. The $\Lambda$ potential depths in both SNM and PNM increase with density. In our future work, these properties should be further constrained using data from $\Lambda$-hypernuclei~\citep{Hashimoto:2006aw,Gal:2016boi}.

Note that we can also fix the binding energy per baryon for pure $\Lambda$-hyperonic matter $E_{\Lambda}(n_{\rm b})=E_0(n_{\rm b})-E_{\rm U}(n_{\rm b})$ by taking $\delta=\delta_{\rm U}=0$. Nevertheless, the pure-\(\Lambda\) limit is governed by the \(\Lambda\Lambda\) sector, for which no independent constraint from double-\(\Lambda\) hypernuclei was included in our analysis. The NAGARA event of ${}^6_{\Lambda\Lambda}$He probes $\Lambda$ matter at rather small densities ($n_\Lambda\lesssim 0.023\ \mathrm{fm}^{-3}$) with the $\Lambda$$\Lambda$ interaction energy $\Delta B_{\Lambda\Lambda} = 0.67 \pm 0.17$ MeV~\cite{Ahn2013_PRC88-014003}. Within the quadratic form adopted in Eq.~(\ref{Eq:Et}), the same coefficient \(E_{\rm U}\) that is constrained by the single-\(\Lambda\) potential also fixes the energy at the pure-\(\Lambda\) endpoint. Our constraints indicate $E_{\Lambda}(n_\Lambda) = -4.889^{+0.188}_{-0.164}$ MeV by setting $n_\Lambda = 0.023\ \mathrm{fm}^{-3}$, suggesting larger $\Lambda$ binding energy than the $\Lambda$$\Lambda$ interaction energy per hyperon $\Delta B_{\Lambda\Lambda}/2$. This should be regarded as an extrapolation of the adopted phenomenological ansatz rather than a reliable prediction, while a realistic extension toward large \(\Lambda\) fractions requires an additional term that independently characterizes the \(\Lambda\Lambda\) sector and can be constrained by double-\(\Lambda\) hypernuclear data, which will be considered in an extended formulation of the model.

\section{\label{sec:con}Summary}
In this work, we introduced the concept of U-spin symmetry energy $E_\mathrm{U}(n_\mathrm{b})$ as an analog to the well-established nuclear symmetry energy $E_\mathrm{sym}(n_\mathrm{b})$, which characterizes the variation in binding energy with the inclusion of hyperons in dense matter. Focusing on the lightest hyperon ($\Lambda$) and neglecting contributions from other hyperons, we then constrained $E_\mathrm{U}(n_\mathrm{b})$ using state-of-the-art nuclear physics data and astrophysical observations via a Bayesian inference approach.

Our analysis indicates that $E_{\rm U}(n_{\rm b})$ is significantly smaller than $9E_{\rm sym}(n_{\rm b})$, indicating that our binding energy functional has a substantially weaker dependence on the U-spin asymmetry than on the corresponding I-spin asymmetry. The effective \(\Lambda\) potential becomes progressively less attractive with increasing density and eventually turns repulsive at high densities. This finding has profound implications for the composition of matter inside neutron stars:
informed by nuclear physics data and astrophysical observations, there is a 55.66\% probability that $\Lambda$ hyperons appear in the EOSs within the causality-satisfying density range. In cases where hyperons emerge, the onset density $n_{\rm b}^{\Lambda}$ typically ranges from $2\,n_0$ to $5\,n_0$, corresponding to neutron stars with masses greater than $1.0\,{\rm M}_\odot$. Furthermore, as the density increases, $E_{\rm U}$ tends toward negative values, implying the disappearance of hyperons in high-density regions. This feature allows the EOSs to support massive neutron stars. The results presented here provide important insights into the EOSs of dense matter and offer a viable resolution to the long-standing ``hyperon puzzle".

\begin{acknowledgements}
The authors would like to thank Prof. Makoto Oka, Prof. Wen-Jie Xie, Dr. Wen-Li Yuan, and Prof. Yingxun Zhang for fruitful discussions. The analysis in this work is inspired by the CompactObject package developed by Huang et al.~\cite{Huang:2024rfg}.
This work was supported by the National Natural Science Foundation of China (Grant No. 12275234), the National SKA Program of China (Grants No. 2020SKA0120300 and No. 2020SKA0120100), and the National Undergraduate Training Program for Innovation and Entrepreneurship and Student Research Training Program (Grant No.~X202511117098). The work of S.H. was supported by Startup Funds from the T.D. Lee Institute and Shanghai Jiao Tong University.
\end{acknowledgements}

\appendix

\section{Nuclear and astrophysical constraints}\label{appA}
The properties of SNM around $n_0$ are well constrained~\citep{Li:2013ola,Oertel:2016bki,Centelles:2008vu,Brown:2013mga}, we thus take $E_0(n_\mathrm{b}) = 0$, $-14.1$ and $-16$ MeV at $n_\mathrm{b}$ = 0, $n_{\mathrm{on}}$, and $n_0$, respectively. Consequently, the skewness ($J_0$) and kurtosis ($Z_0$) parameters are fixed by $J_0 = 33K_0-7783.68$ MeV and $Z_0 = 288K_0-62300.16$ MeV at given $K_0$ (in MeV)~\citep{Oertel:2016bki}. Similarly, we take $E_{\mathrm{sym}}(n_\mathrm{b}) = 0$ and 25.5 MeV at $n_\mathrm{b} =0$ and $n_\mathrm{on}$ according to constraints from finite nuclei~\citep{Centelles:2008vu,Brown:2013mga}. At fixed symmetry energy $E_\mathrm{sym}(n_0)$ and its slope $L_\mathrm{sym}$, the curvature $K_{\mathrm{sym}}$ and skewness $J_{\mathrm{sym}}$ of symmetry energy are determined as $K_{\mathrm{sym}} = 22 L_{\mathrm{sym}} - 194 E_{\mathrm{sym}}(n_0) + 5222$  MeV and $J_{\mathrm{sym}} =144 L_{\mathrm{sym}} - 1584 E_{\mathrm{sym}}(n_0) + 47001.6$  MeV, respectively~\citep{Centelles:2008vu,Brown:2013mga}.
For the U-spin symmetry energy, according to Eq.~(\ref{Eq:VL}), we take $E_\mathrm{U}(n_0) = 5.25$ MeV to reproduce the $\Lambda$ potential depth in SNM at $n_\mathrm{b}=n_0$, i.e., $V_\Lambda(n_0) = E_0(n_0)-8E_\mathrm{U}(n_0)/3= -30$~MeV~\citep{Sun:2017fnf,Rong:2021bim}. By fixing $E_{\mathrm{U}}(n_\mathrm{b}) = 0$ at $n_{\mathrm{b}} = 0$, the skewness parameter $J_{\mathrm{U}} = 9K_{\mathrm{U}} - 54L_{\mathrm{U}} + 850.5$ MeV is determined at given slope $L_{\mathrm{U}}$ and curvature $K_{\mathrm{U}}$ of U-spin symmetry energy. For the Taylor-expansion at saturation density $n_0$, we are now left with
5
free parameters, i.e., $K_0$, $E_{\mathrm{sym}}(n_0)$, $L_{\mathrm{sym}}$, $L_\mathrm{U}$ and $K_\mathrm{U}$.

\begin{table}[t]
\centering
\footnotesize
\setlength{\tabcolsep}{2.5pt}

\caption{ Empirical pressure ($68\%$ CI) of SNM, where $x = n_{\mathrm{b}}/n_0$. The top section ($1.02\leq x \leq2.13$) represents constraints based on chiral NN and 3N interactions~\citep{Drischler:2020hwi,Drischler:2020yad,Drischler:2017wtt}. The middle ($1.30\leq x \leq2.20$) and bottom ($2.0\leq x \leq 4.5$) sections represent results from transport model analyses of kaon production~\citep{Fuchs:2005zg,Lynch:2009vc} and collective flow in HICs~\citep{Danielewicz:2002pu}, respectively.
}
\label{Tab:HIC}
\begin{tabular*}{\columnwidth}{@{\extracolsep{\fill}}cc|cc|cc}
\hline\hline
$x$ & $P$ & $x$ & $P$ & $x$ & $P$ \\
 & (MeV fm$^{-3}$) & & (MeV fm$^{-3}$) & & (MeV fm$^{-3}$) \\
\hline

1.02 & $0.46\pm0.16$ & 1.44 & $3.34\pm2.62$ & 1.81 & $9.82\pm5.80$ \\
1.11 & $0.80\pm0.50$ & 1.46 & $3.61\pm2.64$ & 1.87 & $11.18\pm6.42$ \\
1.22 & $1.38\pm1.08$ & 1.50 & $4.16\pm2.96$ & 1.95 & $13.17\pm7.62$ \\
1.29 & $1.82\pm1.52$ & 1.53 & $4.64\pm3.19$ & 2.02 & $14.77\pm8.37$ \\
1.36 & $2.42\pm2.12$ & 1.59 & $5.45\pm3.59$ & 2.07 & $16.40\pm9.36$ \\
1.37 & $2.52\pm2.18$ & 1.62 & $6.17\pm3.89$ & 2.13 & $17.90\pm10.43$ \\
1.39 & $2.67\pm2.20$ & 1.68 & $7.06\pm4.27$ &      &              \\
1.41 & $2.97\pm2.38$ & 1.74 & $8.44\pm5.11$ &      &              \\

\hline

1.30 & $1.50\pm0.33$ & 1.70 & $5.70\pm1.40$ & 2.10 & $12.85\pm3.70$ \\
1.40 & $2.35\pm0.57$ & 1.80 & $7.10\pm1.87$ & 2.20 & $15.25\pm4.43$ \\
1.50 & $3.10\pm0.67$ & 1.90 & $8.80\pm2.33$ &      &              \\
1.60 & $4.45\pm1.10$ & 2.00 & $10.75\pm2.90$ &      &              \\

\hline

2.0 & $10.40\pm2.00$ & 2.9 & $45.00\pm14.13$ & 3.8 & $95.60\pm30.73$ \\
2.1 & $12.50\pm2.67$ & 3.0 & $50.30\pm16.20$ & 3.9 & $101.85\pm32.83$ \\
2.2 & $14.85\pm3.30$ & 3.1 & $55.65\pm18.03$ & 4.0 & $107.60\pm35.00$ \\
2.3 & $17.75\pm4.17$ & 3.2 & $60.75\pm19.50$ & 4.1 & $112.55\pm37.37$ \\
2.4 & $21.85\pm5.50$ & 3.3 & $65.90\pm20.93$ & 4.2 & $116.85\pm39.97$ \\
2.5 & $26.90\pm7.47$ & 3.4 & $71.25\pm22.57$ & 4.3 & $120.90\pm42.73$ \\
2.6 & $31.85\pm9.37$ & 3.5 & $76.95\pm24.43$ & 4.4 & $125.05\pm45.70$ \\
2.7 & $36.30\pm11.00$ & 3.6 & $82.95\pm26.50$ & 4.5 & $129.55\pm48.70$ \\
2.8 & $40.45\pm12.43$ & 3.7 & $89.25\pm28.63$ &     &              \\

\hline\hline
\end{tabular*}
\end{table}

\begin{table}[t]
\centering
\footnotesize
\setlength{\tabcolsep}{2.5pt}
\caption{\label{Tab:sym} Empirical symmetry energy ($68\%$ CI) as a function of $x=n_{\mathrm{b}}/n_0$. The top section ($0.30\leq x \leq 1.00$) represents the analysis results from HICs~\citep{Tsang:2008fd}, while the bottom section ($0.50\leq x \leq 2.14$) corresponds to $\chi$EFT calculations based on the Gaussian-process framework developed by the BUQEYE Collaboration~\citep{Drischler:2020hwi,Drischler:2020yad,Drischler:2017wtt}.}

\begin{tabular*}{\columnwidth}{@{\extracolsep{\fill}}cc|cc|cc}
\hline\hline

$x$ & $E_{\mathrm{sym}}$ &
$x$ & $E_{\mathrm{sym}}$ &
$x$ & $E_{\mathrm{sym}}$ \\

& MeV & & MeV & & MeV \\

\hline

0.30 & $13.86 \pm 3.74$ & 0.53 & $20.46 \pm 3.53$ & 0.82 & $27.64 \pm 2.59$ \\
0.32 & $14.32 \pm 3.74$ & 0.56 & $21.16 \pm 3.44$ & 0.83 & $28.10 \pm 2.52$ \\
0.33 & $14.68 \pm 3.71$ & 0.58 & $21.74 \pm 3.35$ & 0.86 & $28.66 \pm 2.45$ \\
0.35 & $15.40 \pm 3.76$ & 0.60 & $22.35 \pm 3.32$ & 0.88 & $29.19 \pm 2.37$ \\
0.36 & $15.75 \pm 3.71$ & 0.63 & $23.06 \pm 3.24$ & 0.90 & $29.68 \pm 2.19$ \\
0.38 & $16.22 \pm 3.73$ & 0.65 & $23.63 \pm 3.16$ & 0.92 & $30.17 \pm 2.13$ \\
0.40 & $16.72 \pm 3.65$ & 0.68 & $24.27 \pm 3.07$ & 0.94 & $30.75 \pm 2.07$ \\
0.42 & $17.27 \pm 3.65$ & 0.71 & $24.90 \pm 2.98$ & 0.96 & $31.23 \pm 1.92$ \\
0.43 & $17.65 \pm 3.67$ & 0.73 & $25.51 \pm 2.92$ & 0.98 & $31.64 \pm 1.93$ \\
0.45 & $18.26 \pm 3.60$ & 0.75 & $26.02 \pm 2.84$ & 1.00 & $31.98 \pm 1.81$ \\
0.48 & $19.02 \pm 3.54$ & 0.78 & $26.68 \pm 2.75$ &      &                  \\
0.50 & $19.78 \pm 3.51$ & 0.80 & $27.17 \pm 2.75$ &      &                  \\

\hline

0.50 & $20.10 \pm 1.13$ & 1.12 & $32.89 \pm 4.23$ & 1.87 & $43.61 \pm 12.89$ \\
0.57 & $21.44 \pm 1.24$ & 1.27 & $35.26 \pm 5.36$ & 1.94 & $44.02 \pm 14.33$ \\
0.64 & $23.09 \pm 1.44$ & 1.41 & $37.63 \pm 6.70$ & 2.00 & $44.43 \pm 15.15$ \\
0.73 & $25.05 \pm 1.96$ & 1.54 & $39.69 \pm 8.35$ & 2.08 & $45.05 \pm 17.01$ \\
0.87 & $27.94 \pm 2.58$ & 1.67 & $41.44 \pm 9.90$ & 2.14 & $44.95 \pm 18.14$ \\
0.98 & $30.31 \pm 3.09$ & 1.79 & $42.78 \pm 11.86$ &      &                  \\

\hline\hline
\end{tabular*}
\end{table}

\begin{table}[t]
\centering
\footnotesize
\setlength{\tabcolsep}{2.5pt}
\caption{\label{Tab:U-lambda}The empirical $\Lambda$ hyperon potential well depth in SNM (68$\%$ CI) versus $x = n_{\mathrm{b}}/n_0$, based on SU(3) $\chi$EFT calculations incorporating three-body forces~\citep{Gerstung:2020ktv}.}
\begin{tabular*}{\columnwidth}{@{\extracolsep{\fill}}cc|cc|cc}
\hline\hline
$x$ & $V_{\Lambda}$ & $x$ & $V_{\Lambda}$ & $x$ & $V_{\Lambda}$ \\
    & MeV & & MeV & & MeV \\
\hline
1.20 & $-28.39 \pm 0.43$ & 1.82 & $-8.04 \pm 3.50$  & 2.45 & $30.02 \pm 9.28$  \\
1.24 & $-27.77 \pm 0.54$ & 1.86 & $-6.24 \pm 3.76$  & 2.49 & $32.72 \pm 9.71$  \\
1.27 & $-27.07 \pm 0.66$ & 1.90 & $-4.39 \pm 4.03$  & 2.52 & $35.46 \pm 10.16$ \\
1.31 & $-26.30 \pm 0.79$ & 1.93 & $-2.48 \pm 4.31$  & 2.56 & $38.25 \pm 10.61$ \\
1.35 & $-25.44 \pm 0.93$ & 1.97 & $-0.51 \pm 4.59$  & 2.60 & $41.08 \pm 11.08$ \\
1.38 & $-24.52 \pm 1.08$ & 2.01 & $1.52 \pm 4.89$   & 2.63 & $43.95 \pm 11.56$ \\
1.42 & $-23.52 \pm 1.23$ & 2.04 & $3.61 \pm 5.20$   & 2.67 & $46.87 \pm 12.05$ \\
1.46 & $-22.45 \pm 1.39$ & 2.08 & $5.75 \pm 5.52$   & 2.71 & $49.83 \pm 12.55$ \\
1.49 & $-21.31 \pm 1.57$ & 2.12 & $7.94 \pm 5.85$   & 2.74 & $52.83 \pm 13.06$ \\
1.53 & $-20.10 \pm 1.74$ & 2.16 & $10.19 \pm 6.19$  & 2.78 & $55.87 \pm 13.58$ \\
1.57 & $-18.83 \pm 1.93$ & 2.19 & $12.49 \pm 6.54$  & 2.82 & $58.95 \pm 14.12$ \\
1.60 & $-17.48 \pm 2.13$ & 2.23 & $14.85 \pm 6.90$  & 2.85 & $62.07 \pm 14.67$ \\
1.64 & $-16.07 \pm 2.33$ & 2.27 & $17.25 \pm 7.27$  & 2.89 & $65.22 \pm 15.22$ \\
1.68 & $-14.59 \pm 2.55$ & 2.30 & $19.71 \pm 7.65$  & 2.93 & $68.41 \pm 15.79$ \\
1.71 & $-13.05 \pm 2.77$ & 2.34 & $22.22 \pm 8.04$  & 2.96 & $71.64 \pm 16.37$ \\
1.75 & $-11.44 \pm 3.00$ & 2.38 & $24.77 \pm 8.44$  & 3.00 & $74.91 \pm 16.97$ \\
1.79 & $-9.77 \pm 3.24$  & 2.41 & $27.37 \pm 8.85$  &      &                   \\
\hline\hline
\end{tabular*}
\end{table}

\begin{table}[t]
\centering
\caption{\label{Tab:data_astro}
Astrophysical constraints (68\% CL) on the masses and radii of compact stars adopted in the present work.
}

\begin{tabular}{lccc}
\hline\hline
Source  & Mass ($M_\odot$) & Radius (km) \\
\hline

GW170817~\cite{LIGOScientific:2018cki} & $1.36^{+0.14}_{-0.13}$
& $11.88^{+0.98}_{-0.98}$ \\

PSR J0030+0451~\cite{Riley:2019yda}
& $1.34^{+0.15}_{-0.15}$
& $12.71^{+1.13}_{-1.18}$ \\

PSR J0740+6620~\cite{Miller:2021qha}
& $2.07^{+0.07}_{-0.07}$
& $12.39^{+1.30}_{-0.98}$ \\

PSR J0437$-$4715~\cite{Choudhury:2024xbk}
& $1.42^{+0.04}_{-0.04}$
& $11.36^{+0.94}_{-0.62}$ \\

PSR J0614$-$3329~\cite{Mauviard:2025dmd}
& $1.44^{+0.06}_{-0.07}$
& $10.29^{+1.01}_{-0.86}$ \\

HESS J1731$-$347~\cite{Doroshenko:2022nwp}
& $0.77^{+0.20}_{-0.17}$
& $10.4^{+0.86}_{-0.78}$ \\

\hline\hline
\end{tabular}
\end{table}

Regarding the nuclear physics constraints, beside the constraints on nuclear saturation properties, we consider the HIC data and predictions from $\chi$EFT calculations. As indicated in Table~\ref{Tab:HIC}, the SNM pressure is constrained by empirical data from heavy-ion collisions, i.e., transport model analyses of kaon production~\citep{Fuchs:2005zg,Lynch:2009vc} and collective flow~\citep{Danielewicz:2002pu}. Additionally, theoretical constraints on the SNM pressure from $\chi$EFT are also incorporated~\citep{Drischler:2020hwi}. In Table \ref{Tab:sym}, regarding the symmetry energy, we consider not only the isospin diffusion data from transport models in HIC~\citep{Tsang:2008fd} but also the theoretical calculations from $\chi$EFT based on the Gaussian process framework developed by the BUQEYE Collaboration~\citep{Drischler:2020hwi,Drischler:2020yad,Drischler:2017wtt}. Furthermore, to incorporate hyperonic degrees of freedom, Table \ref{Tab:U-lambda} lists the empirical $\Lambda$ potential well depth in SNM. These values are derived from SU(3) $\chi$EFT including essential three-body forces~\citep{Gerstung:2020ktv}, providing a reasonable benchmark for the hyperon-nucleon interaction at supra-saturation densities.

For the astrophysical constraints, we consider the following mass and radius measurements of neutron stars, i.e., the binary neutron star merger event {GRB} 170817A-{GW}170817-{AT} 2017gfo~\citep{LIGOScientific:2018cki}, the pulse-profile modelings with the {NICER} and {XMM}-Newton data for PSR J0030+0451,  PSR J0740+6620, PSR J0437-4715, and PSR J0614-3329~\citep{Riley:2019yda,Riley:2021pdl,Miller:2019cac,Miller:2021qha,Choudhury:2024xbk,Mauviard:2025dmd}, as well as the central compact object (CCO) within the supernova remnant HESS J1731-347~\citep{Doroshenko:2022nwp}. The corresponding constraints are then summarized in Table~\ref{Tab:data_astro}.

\section{\label{sec:bayes}Bayesian inference approach}

The Bayesian inference approach is formalized through Bayes' theorem with the probability updated via new data, where in this work we adopt the framework proposed in Refs.~\citep{Greif:2018njt,Raaijmakers:2019qny,Raaijmakers:2019dks,Huang:2023grj}. The posterior distribution of parameter set $\boldsymbol{\theta}$ is the product of the corresponding prior distribution and the nuisance-marginalized likelihood function, i.e.,
\begin{equation}
 p(\boldsymbol{\theta} \mid \boldsymbol{d}, \mathcal{M}) \propto p(\boldsymbol{\theta} \mid \mathcal{M})  p(\boldsymbol{d} \mid \boldsymbol{\theta}, \mathcal{M}).
\label{Eq:Bay1}
\end{equation}
where  $\mathcal{M}$ denotes the model and $\boldsymbol{d}$ the dataset. The affine-invariant ensemble MCMC sampler implemented in the \texttt{emcee} package~\citep{Foreman-Mackey:2012any} is employed to sample the parameter vector \(\boldsymbol{\theta}\). We use 100 walkers and 10,000 steps for each walker, resulting in \(10^6\) raw samples. The first 2,000 steps of each walker are discarded as burn-in. After the burn-in stage, the individual chains become stationary. For the posterior predictions presented in this work, 10,000 parameter sets are randomly selected from the retained samples to improve computational efficiency.

\begin{table}[t] \centering
\caption{\label{Tab:prior}Prior ranges of the parameters used in this work, where $\mathcal{U}$ and $\mathcal{N}$ indicate Uniform and Gaussian distributions, respectively.}
\renewcommand{\arraystretch}{1.2} %
\begin{tabular*}{\linewidth}{lll} 
Segment & Parameters & Prior Distribution \\
\hline
Seg.~0 & $K_0$ / MeV& $\mathcal{N}(240,20)$ \\
      & $E_{\mathrm{sym}} (n_0)$ / MeV& $\mathcal{U}(28,36)$ \\
      & $L_{\mathrm{sym}}$ / MeV& $\mathcal{U}(-100,200)$ \\
      & $L_{\mathrm{U}}$/ MeV& $\mathcal{U}(-100,100)$ \\
      & $K_{\mathrm{U}}$ / MeV& $\mathcal{U}(-200,200)$ \\
\hline
Seg.~1 & $n_1$ / fm$^{-3}$& $0.24$ \\
      & $J_{\mathrm{0,1}}$/ MeV & $\mathcal{U} (-5 \cdot 10^4, 5 \cdot 10^4)$ \\
      & $J_{\mathrm{sym,1}}$/ MeV & $\mathcal{U} (-2 \cdot 10^5, 2 \cdot 10^5)$ \\
      & $J_{\mathrm{U,1}}$/ MeV & $\mathcal{U} (-2 \cdot 10^5, 2 \cdot 10^5)$ \\

\hline& $n_2$ / fm$^{-3}$ & $\mathcal{U}(n_1, 3n_1)$ \\
Seg.~2 & $J_{\mathrm{0,2}}$ / MeV& $\mathcal{U} (-2 \cdot 10^6, 2 \cdot 10^6)$ \\
      & $J_{\mathrm{sym,2}}$ / MeV& $\mathcal{U} (-2 \cdot 10^6, 2 \cdot 10^6)$ \\
      & $J_{\mathrm{U,2}}$ / MeV & $\mathcal{U} (-2 \cdot 10^6, 2 \cdot 10^6)$ \\

\hline& $n_3$ / fm$^{-3}$& $\mathcal{U}(n_2, 3n_2)$ \\
Seg.~3 & $J_{\mathrm{0,3}}$ / MeV& $\mathcal{U} (-3 \cdot 10^6, 3 \cdot 10^6)$ \\
      & $J_{\mathrm{sym,3}}$ / MeV& $\mathcal{U} (-3 \cdot 10^6, 3 \cdot 10^6)$ \\
      & $J_{\mathrm{U,3}}$ / MeV& $\mathcal{U} (-3 \cdot 10^6, 3 \cdot 10^6)$ \\

\hline& $n_4$ / fm$^{-3}$& $\mathcal{U}(n_3, 3n_3)$ \\
Seg.~4 & $J_{\mathrm{0,4}}$ / MeV& $\mathcal{U} (-4 \cdot 10^6, 4 \cdot 10^6)$ \\
      & $J_{\mathrm{sym,4}}$ / MeV& $\mathcal{U} (-4 \cdot 10^6, 4 \cdot 10^6)$ \\
      & $J_{\mathrm{U,4}}$ / MeV& $\mathcal{U} (-4 \cdot 10^6, 4 \cdot 10^6)$ \\

\hline& $n_{5}$ / fm$^{-3}$ & $\mathcal{U}(n_4, 1.5)$ \\
Seg.~5 & $J_{\mathrm{0,5}}$ / MeV& $\mathcal{U} (-5 \cdot 10^6, 5 \cdot 10^6)$ \\
      & $J_{\mathrm{sym,5}}$ / MeV& $\mathcal{U} (-5 \cdot 10^6, 5 \cdot 10^6)$ \\
      & $J_{\mathrm{U,5}}$ / MeV& $\mathcal{U} (-5 \cdot 10^6, 5 \cdot 10^6)$ \\
\hline\hline
\end{tabular*}
\end{table}

As indicated in Sec.~\ref{sec:the}, our EOS model is divided into several segments below $n_{\mathrm{b}} = 1.5$ $\text{fm}^{-3}$. The number of segments, ranging from 1 to 7, was tested for both normal nuclear matter and hyperonic matter. In the low-density region, due to the relatively abundant constraints from nuclear
experiments, $\chi$EFT calculations and transport model of intermediate-energy heavy-ion collisions (HIC), the results turn out to be quite insensitive to the number of density segments. In the high-density region, however, there exist evident fluctuations while both the posterior distributions of parameters and EOSs converge as the number of segments increases. Considering computational efficiency and accuracy, our EOS model divides $E_0$, $E_{\mathrm{sym}}$, and $E_{\mathrm{U}}$ into six segments. Consequently, for hyperonic matter, our EOS model incorporates 24 free parameters, forming a 24-dimensional parameter vector $\boldsymbol{\theta}$. While the incompressibility parameter $K_0$ is sampled from a Gaussian distribution~\citep{Xie:2020kta}, the remaining parameters $\theta_{i=1,\dots,23}$ are sampled uniformly within the prior ranges specified in Table~\ref{Tab:prior}. Note that if the U-spin symmetry energy is not explicitly accounted for, the number of free parameters in the EOS model reduces to 17.

Finally, by including these constraints the posterior distribution of parameter set $\boldsymbol{\theta}$ in Eq.~(\ref{Eq:Bay1}) can be recast as
\begin{equation}
\begin{aligned}
&p(\boldsymbol{\theta} \mid \boldsymbol{d}, \mathcal{M}) \propto p(\boldsymbol{\theta} \mid \mathcal{M}) \prod_{i} p\left(P_{i} \mid \boldsymbol{d}_{\mathrm{Nucl, \textrm {i}}}\right) \\
&\times \prod_{j}\int  p(\varepsilon_{c,j} \mid \boldsymbol{\theta}, \mathcal{M}) p\left(M_{j}, R_{j} \mid d_{\mathrm{Astro}, \mathrm{j}}\right) \mbox{d}\varepsilon_{c,j},
\end{aligned}
\end{equation}
where $\boldsymbol{\varepsilon}_{c,j}$ represents the central energy densities of neutron stars with sufficiently large prior bounds $p(\varepsilon_{c,j} \mid \boldsymbol{\theta}, \mathcal{M})$. Note that the sampling process incorporates fundamental physical constraints, i.e., all EOSs should meet the two-solar-mass constraint $M_{\mathrm{max}} > 1.97 \,{\rm M}_\odot$, be thermodynamically stable and causal ($0< {\mbox{d}P}/{\mbox{d}\varepsilon} < 1$) at $\varepsilon\leq \varepsilon_{c,1.97 \, {\rm M}_\odot}$.


\begin{thebibliography}{100}
\expandafter\ifx\csname url\endcsname\relax
  \def\url#1{\texttt{#1}}\fi
\expandafter\ifx\csname urlprefix\endcsname\relax\def\urlprefix{URL }\fi
\expandafter\ifx\csname href\endcsname\relax
  \def\href#1#2{#2} \def\path#1{#1}\fi

\bibitem{Greiner1994}
W.~Greiner, B.~M{\"u}ller, Quantum Mechanics: Symmetries, Springer Science \&
  Business Media, 1994.
\newblock \href {https://doi.org/10.1007/978-3-642-57976-9}
  {\path{doi:10.1007/978-3-642-57976-9}}.

\bibitem{Darewych:1983yw}
J.~W. Darewych, M.~Horbatsch, R.~Koniuk, {The Photon Decays of Baryons With
  Strangeness}, Phys. Rev. D 28 (1983) 1125.
\newblock \href {http://dx.doi.org/10.1103/PhysRevD.28.1125}
  {\path{doi:10.1103/PhysRevD.28.1125}}.

\bibitem{Julia-Diaz:2007mae}
B.~Julia-Diaz, T.~S.~H. Lee, A.~Matsuyama, T.~Sato, L.~C. Smith, {Dynamical
  coupled-channels effects on pion photoproduction}, Phys. Rev. C 77 (2008)
  045205.
\newblock \href {http://arxiv.org/abs/0712.2283} {\path{arXiv:0712.2283}},
  \href {http://dx.doi.org/10.1103/PhysRevC.77.045205}
  {\path{doi:10.1103/PhysRevC.77.045205}}.

\bibitem{Jia:2019zxi}
C.-P. Jia, D.~Wang, F.-S. Yu, {Charmed baryon decays in $SU(3)_F$ symmetry},
  Nucl. Phys. B 956 (2020) 115048.
\newblock \href {http://arxiv.org/abs/1910.00876} {\path{arXiv:1910.00876}},
  \href {http://dx.doi.org/10.1016/j.nuclphysb.2020.115048}
  {\path{doi:10.1016/j.nuclphysb.2020.115048}}.

\bibitem{Zhang:2025jnw}
B.-n. Zhang, D.~Wang, {U-spin conjugate CP violation relations in bottom baryon
  decays}, Phys. Lett. B 868 (2025) 139674.
\newblock \href {http://arxiv.org/abs/2503.21885} {\path{arXiv:2503.21885}},
  \href {http://dx.doi.org/10.1016/j.physletb.2025.139674}
  {\path{doi:10.1016/j.physletb.2025.139674}}.

\bibitem{Centelles:2008vu}
M.~Centelles, X.~Roca-Maza, X.~Vinas, M.~Warda, {Nuclear symmetry energy probed
  by neutron skin thickness of nuclei}, Phys. Rev. Lett. 102 (2009) 122502.
\newblock \href {http://arxiv.org/abs/0806.2886} {\path{arXiv:0806.2886}},
  \href {http://dx.doi.org/10.1103/PhysRevLett.102.122502}
  {\path{doi:10.1103/PhysRevLett.102.122502}}.

\bibitem{Brown:2013mga}
B.~A. Brown, {Constraints on the Skyrme Equations of State from Properties of
  Doubly Magic Nuclei}, Phys. Rev. Lett. 111~(23) (2013) 232502.
\newblock \href {http://arxiv.org/abs/1308.3664} {\path{arXiv:1308.3664}},
  \href {http://dx.doi.org/10.1103/PhysRevLett.111.232502}
  {\path{doi:10.1103/PhysRevLett.111.232502}}.

\bibitem{Li:2013ola}
B.-A. Li, X.~Han, {Constraining the neutron-proton effective mass splitting
  using empirical constraints on the density dependence of nuclear symmetry
  energy around normal density}, Phys. Lett. B 727 (2013) 276--281.
\newblock \href {http://arxiv.org/abs/1304.3368} {\path{arXiv:1304.3368}},
  \href {http://dx.doi.org/10.1016/j.physletb.2013.10.006}
  {\path{doi:10.1016/j.physletb.2013.10.006}}.

\bibitem{Oertel:2016bki}
M.~Oertel, M.~Hempel, T.~Kl{\"a}hn, S.~Typel, {Equations of state for
  supernovae and compact stars}, Rev. Mod. Phys. 89~(1) (2017) 015007.
\newblock \href {http://arxiv.org/abs/1610.03361} {\path{arXiv:1610.03361}},
  \href {http://dx.doi.org/10.1103/RevModPhys.89.015007}
  {\path{doi:10.1103/RevModPhys.89.015007}}.

\bibitem{Hashimoto:2006aw}
O.~Hashimoto, H.~Tamura, {Spectroscopy of Lambda hypernuclei}, Prog. Part.
  Nucl. Phys. 57 (2006) 564--653.
\newblock \href {http://dx.doi.org/10.1016/j.ppnp.2005.07.001}
  {\path{doi:10.1016/j.ppnp.2005.07.001}}.

\bibitem{KEKE176:2009jzw}
S.~Aoki, et~al., {Nuclear capture at rest of $\Xi^-$ hyperons}, Nucl. Phys. A
  828 (2009) 191--232.
\newblock \href {http://dx.doi.org/10.1016/j.nuclphysa.2009.07.005}
  {\path{doi:10.1016/j.nuclphysa.2009.07.005}}.

\bibitem{E373KEK-PS:2013dfg}
J.~K. Ahn, et~al., {Double-$\Lambda$ hypernuclei observed in a hybrid emulsion
  experiment}, Phys. Rev. C 88~(1) (2013) 014003.
\newblock \href {http://dx.doi.org/10.1103/PhysRevC.88.014003}
  {\path{doi:10.1103/PhysRevC.88.014003}}.

\bibitem{Gal:2016boi}
A.~Gal, E.~V. Hungerford, D.~J. Millener, {Strangeness in nuclear physics},
  Rev. Mod. Phys. 88~(3) (2016) 035004.
\newblock \href {http://arxiv.org/abs/1605.00557} {\path{arXiv:1605.00557}},
  \href {http://dx.doi.org/10.1103/RevModPhys.88.035004}
  {\path{doi:10.1103/RevModPhys.88.035004}}.

\bibitem{Gal:1971gb}
A.~Gal, J.~M. Soper, R.~H. Dalitz, {Shell-model analysis of lambda binding
  energies for p-shell hypernuclei. 1. Basic formulas and matrix elements for
  lambda N and lambda NN forces}, Annals Phys. 63 (1971) 53--126.
\newblock \href {http://dx.doi.org/10.1016/0003-4916(71)90297-1}
  {\path{doi:10.1016/0003-4916(71)90297-1}}.

\bibitem{Dalitz:1978hw}
R.~H. Dalitz, A.~Gal, {The formation of, and the gamma-radiation from, the
  p-shell hypernuclei}, Annals Phys. 116 (1978) 167--243.
\newblock \href {http://dx.doi.org/10.1016/0003-4916(78)90008-8}
  {\path{doi:10.1016/0003-4916(78)90008-8}}.

\bibitem{Millener:2008zz}
D.~J. Millener, {Shell-model interpretation of gamma-ray transitions in p-shell
  hypernuclei}, Nucl. Phys. A 804 (2008) 84--98.
\newblock \href {http://dx.doi.org/10.1016/j.nuclphysa.2008.02.252}
  {\path{doi:10.1016/j.nuclphysa.2008.02.252}}.

\bibitem{Millener:2013mwa}
D.~J. Millener, {Shell-model calculations applied to the $\gamma$-ray
  spectroscopy of light $\Lambda$ hypernuclei}, Nucl. Phys. A 914 (2013)
  109--118.
\newblock \href {http://dx.doi.org/10.1016/j.nuclphysa.2013.01.023}
  {\path{doi:10.1016/j.nuclphysa.2013.01.023}}.

\bibitem{Motoba:1983kbv}
T.~Motoba, H.~Bando, K.~Ikeda, {LIGHT P SHELL LAMBDA HYPERNUCLEI BY THE
  MICROSCOPIC THREE CLUSTER MODEL}, Prog. Theor. Phys. 70 (1983) 189--221.
\newblock \href {http://dx.doi.org/10.1143/PTP.70.189}
  {\path{doi:10.1143/PTP.70.189}}.

\bibitem{Hiyama:2009ki}
E.~Hiyama, Y.~Yamamoto, T.~Motoba, M.~Kamimura, {Structure of A=7 iso-triplet
  $\Lambda$ hypernuclei studied with he four-body model}, Phys. Rev. C 80
  (2009) 054321.
\newblock \href {http://arxiv.org/abs/0911.4013} {\path{arXiv:0911.4013}},
  \href {http://dx.doi.org/10.1103/PhysRevC.80.054321}
  {\path{doi:10.1103/PhysRevC.80.054321}}.

\bibitem{Bando:1990yi}
H.~Bando, T.~Motoba, J.~Zofka, {Production, structure and decay of
  hypernuclei}, Int. J. Mod. Phys. A 5 (1990) 4021--4198.
\newblock \href {http://dx.doi.org/10.1142/S0217751X90001732}
  {\path{doi:10.1142/S0217751X90001732}}.

\bibitem{Isaka:2013sfa}
M.~Isaka, M.~Kimura, A.~Dot{\'e}, A.~Ohnishi, {Splitting of the $p$ orbit in
  triaxially deformed $^{25}_{\Lambda}$Mg}, Phys. Rev. C 87~(2) (2013) 021304.
\newblock \href {http://dx.doi.org/10.1103/PhysRevC.87.021304}
  {\path{doi:10.1103/PhysRevC.87.021304}}.

\bibitem{Hu:2013tma}
J.~N. Hu, A.~Li, H.~Toki, W.~Zuo, {Extended quark mean-field model for neutron
  stars}, Phys. Rev. C 89~(2) (2014) 025802.
\newblock \href {http://arxiv.org/abs/1307.4154} {\path{arXiv:1307.4154}},
  \href {http://dx.doi.org/10.1103/PhysRevC.89.025802}
  {\path{doi:10.1103/PhysRevC.89.025802}}.

\bibitem{Brockmann:1977es}
R.~Brockmann, W.~Weise, {Lambda n Interaction and Spin Orbit Coupling in
  Hypernuclei}, Phys. Lett. B 69 (1977) 167--169.
\newblock \href {http://dx.doi.org/10.1016/0370-2693(77)90635-9}
  {\path{doi:10.1016/0370-2693(77)90635-9}}.

\bibitem{Boguta:1980wi}
J.~Boguta, S.~Bohrmann, {RELATIVISTIC QUANTUM FIELD THEORY OF LAMBDA
  HYPERNUCLEI}, Phys. Lett. B 102 (1981) 93--96.
\newblock \href {http://dx.doi.org/10.1016/0370-2693(81)91037-6}
  {\path{doi:10.1016/0370-2693(81)91037-6}}.

\bibitem{Mares:1993xb}
J.~Mares, B.~K. Jennings, {Relativistic description of Lambda, Sigma and Xi
  hypernuclei}, Phys. Rev. C 49 (1994) 2472--2478.
\newblock \href {http://dx.doi.org/10.1103/PhysRevC.49.2472}
  {\path{doi:10.1103/PhysRevC.49.2472}}.

\bibitem{Sugahara:1994pn}
Y.~Sugahara, H.~Toki, {Relativistic mean field theory for Lambda hypernuclei
  and neutron stars}, Prog. Theor. Phys. 92 (1994) 803--813.
\newblock \href {http://dx.doi.org/10.1143/PTP.92.803}
  {\path{doi:10.1143/PTP.92.803}}.

\bibitem{Song:2010qhk}
C.~Song, J.~Yao, H.~F. Lu, J.~Meng, {Lambda and Anti-Lambda Hypernuclei in
  Relativistic Mean-field Theory}, Int. J. Mod. Phys. E 19 (2010) 2538--2545.
\newblock \href {http://arxiv.org/abs/1004.2395} {\path{arXiv:1004.2395}},
  \href {http://dx.doi.org/10.1142/S0218301310017058}
  {\path{doi:10.1142/S0218301310017058}}.

\bibitem{Tanimura:2011ji}
Y.~Tanimura, K.~Hagino, {Description of single-Lambda hypernuclei with
  relativistic point coupling model}, Phys. Rev. C 85 (2012) 014306.
\newblock \href {http://arxiv.org/abs/1111.1488} {\path{arXiv:1111.1488}},
  \href {http://dx.doi.org/10.1103/PhysRevC.85.014306}
  {\path{doi:10.1103/PhysRevC.85.014306}}.

\bibitem{Liu:2018img}
Z.~X. Liu, C.~J. Xia, W.~L. Lu, Y.~X. Li, J.~N. Hu, T.~T. Sun, {Relativistic
  mean-field approach for {Lambda},{Xi} , and {Sigma} hypernuclei}, Phys. Rev.
  C 98~(2) (2018) 024316.
\newblock \href {http://dx.doi.org/10.1103/PhysRevC.98.024316}
  {\path{doi:10.1103/PhysRevC.98.024316}}.

\bibitem{Rong:2021bim}
Y.-T. Rong, Z.-H. Tu, S.-G. Zhou, {New effective interactions for hypernuclei
  in a density-dependent relativistic mean field model}, Phys. Rev. C 104~(5)
  (2021) 054321.
\newblock \href {http://arxiv.org/abs/2103.10706} {\path{arXiv:2103.10706}},
  \href {http://dx.doi.org/10.1103/PhysRevC.104.054321}
  {\path{doi:10.1103/PhysRevC.104.054321}}.

\bibitem{Rong:2025wnv}
Y.-T. Rong, D.~Yang, C.-J. Xia, T.-T. Sun, {Constraints on $\Lambda N$
  Effective Interactions from Mirror Hypernuclei in a Deformed Relativistic
  Hartree-Bogoliubov Model}\href {http://arxiv.org/abs/2506.13499}
  {\path{arXiv:2506.13499}}.

\bibitem{Zhou:2007zze}
X.-R. Zhou, H.~J. Schulze, H.~Sagawa, C.-X. Wu, E.-G. Zhao, {Hypernuclei in the
  deformed Skyrme-Hartree-Fock approach}, Phys. Rev. C 76 (2007) 034312.
\newblock \href {http://dx.doi.org/10.1103/PhysRevC.76.034312}
  {\path{doi:10.1103/PhysRevC.76.034312}}.

\bibitem{Tsushima:1997cu}
K.~Tsushima, K.~Saito, J.~Haidenbauer, A.~W. Thomas, {The Quark - meson
  coupling model for Lambda, Sigma and Xi hypernuclei}, Nucl. Phys. A 630
  (1998) 691--718.
\newblock \href {http://arxiv.org/abs/nucl-th/9707022}
  {\path{arXiv:nucl-th/9707022}}, \href
  {http://dx.doi.org/10.1016/S0375-9474(98)00806-9}
  {\path{doi:10.1016/S0375-9474(98)00806-9}}.

\bibitem{Guichon:2008zz}
P.~A.~M. Guichon, A.~W. Thomas, K.~Tsushima, {Binding of hypernuclei in the
  latest quark-meson coupling model}, Nucl. Phys. A 814 (2008) 66--73.
\newblock \href {http://arxiv.org/abs/0712.1925} {\path{arXiv:0712.1925}},
  \href {http://dx.doi.org/10.1016/j.nuclphysa.2008.10.001}
  {\path{doi:10.1016/j.nuclphysa.2008.10.001}}.

\bibitem{Sun:2017fnf}
T.-T. Sun, C.-J. Xia, S.-S. Zhang, M.~S. Smith, {Massive neutron stars and
  $\Lambda$-hypernuclei in relativistic mean field models}, Chin. Phys. C
  42~(2) (2018) 025101.
\newblock \href {http://arxiv.org/abs/1712.05569} {\path{arXiv:1712.05569}},
  \href {http://dx.doi.org/10.1088/1674-1137/42/2/025101}
  {\path{doi:10.1088/1674-1137/42/2/025101}}.

\bibitem{Sun:2018tmw}
T.-T. Sun, S.-S. Zhang, Q.-L. Zhang, C.-J. Xia, {Strangeness and $\Delta$
  resonance in compact stars with relativistic-mean-field models}, Phys. Rev. D
  99~(2) (2019) 023004.
\newblock \href {http://arxiv.org/abs/1808.02207} {\path{arXiv:1808.02207}},
  \href {http://dx.doi.org/10.1103/PhysRevD.99.023004}
  {\path{doi:10.1103/PhysRevD.99.023004}}.

\bibitem{Demorest:2010bx}
P.~Demorest, T.~Pennucci, S.~Ransom, M.~Roberts, J.~Hessels, {Shapiro Delay
  Measurement of A Two Solar Mass Neutron Star}, Nature 467 (2010) 1081--1083.
\newblock \href {http://arxiv.org/abs/1010.5788} {\path{arXiv:1010.5788}},
  \href {http://dx.doi.org/10.1038/nature09466}
  {\path{doi:10.1038/nature09466}}.

\bibitem{Antoniadis:2013pzd}
J.~Antoniadis, et~al., {A Massive Pulsar in a Compact Relativistic Binary},
  Science 340 (2013) 6131.
\newblock \href {http://arxiv.org/abs/1304.6875} {\path{arXiv:1304.6875}},
  \href {http://dx.doi.org/10.1126/science.1233232}
  {\path{doi:10.1126/science.1233232}}.

\bibitem{Fonseca:2016tux}
E.~Fonseca, et~al., {The NANOGrav Nine-year Data Set: Mass and Geometric
  Measurements of Binary Millisecond Pulsars}, Astrophys. J. 832~(2) (2016)
  167.
\newblock \href {http://arxiv.org/abs/1603.00545} {\path{arXiv:1603.00545}},
  \href {http://dx.doi.org/10.3847/0004-637X/832/2/167}
  {\path{doi:10.3847/0004-637X/832/2/167}}.

\bibitem{Vidana:2015faa}
I.~Vida{\~n}a, {Hyperons and neutron stars}, AIP Conf. Proc. 1645~(1) (2015)
  79--85.
\newblock \href {http://dx.doi.org/10.1063/1.4909561}
  {\path{doi:10.1063/1.4909561}}.

\bibitem{Weissenborn:2011ut}
S.~Weissenborn, D.~Chatterjee, J.~Schaffner-Bielich, {Hyperons and massive
  neutron stars: vector repulsion and SU(3) symmetry}, Phys. Rev. C 85~(6)
  (2012) 065802, [Erratum: Phys.Rev.C 90, 019904 (2014)].
\newblock \href {http://arxiv.org/abs/1112.0234} {\path{arXiv:1112.0234}},
  \href {http://dx.doi.org/10.1103/PhysRevC.85.065802}
  {\path{doi:10.1103/PhysRevC.85.065802}}.

\bibitem{Bednarek:2011gd}
I.~Bednarek, P.~Haensel, J.~L. Zdunik, M.~Bejger, R.~Manka, {Hyperons in
  neutron-star cores and two-solar-mass pulsar}, Astron. Astrophys. 543 (2012)
  A157.
\newblock \href {http://arxiv.org/abs/1111.6942} {\path{arXiv:1111.6942}},
  \href {http://dx.doi.org/10.1051/0004-6361/201118560}
  {\path{doi:10.1051/0004-6361/201118560}}.

\bibitem{Oertel:2014qza}
M.~Oertel, C.~Provid{\^e}ncia, F.~Gulminelli, A.~R. Raduta, {Hyperons in
  neutron star matter within relativistic mean-field models}, J. Phys. G 42~(7)
  (2015) 075202.
\newblock \href {http://arxiv.org/abs/1412.4545} {\path{arXiv:1412.4545}},
  \href {http://dx.doi.org/10.1088/0954-3899/42/7/075202}
  {\path{doi:10.1088/0954-3899/42/7/075202}}.

\bibitem{Maslov:2015msa}
K.~A. Maslov, E.~E. Kolomeitsev, D.~N. Voskresensky, {Solution of the Hyperon
  Puzzle within a Relativistic Mean-Field Model}, Phys. Lett. B 748 (2015)
  369--375.
\newblock \href {http://arxiv.org/abs/1504.02915} {\path{arXiv:1504.02915}},
  \href {http://dx.doi.org/10.1016/j.physletb.2015.07.032}
  {\path{doi:10.1016/j.physletb.2015.07.032}}.

\bibitem{Maslov:2015wba}
K.~A. Maslov, E.~E. Kolomeitsev, D.~N. Voskresensky, {Relativistic Mean-Field
  Models with Scaled Hadron Masses and Couplings: Hyperons and Maximum Neutron
  Star Mass}, Nucl. Phys. A 950 (2016) 64--109.
\newblock \href {http://arxiv.org/abs/1509.02538} {\path{arXiv:1509.02538}},
  \href {http://dx.doi.org/10.1016/j.nuclphysa.2016.03.011}
  {\path{doi:10.1016/j.nuclphysa.2016.03.011}}.

\bibitem{Takatsuka:2002adv}
T.~Takatsuka, S.~Nishizaki, Y.~Yamamoto, {Necessity of extra repulsion in
  hypernuclear systems: Suggestion from neutron stars}, Eur. Phys. J. A 13~(1)
  (2002) 213--215.
\newblock \href {http://dx.doi.org/10.1140/epja1339-35}
  {\path{doi:10.1140/epja1339-35}}.

\bibitem{Vidana:2010ip}
I.~Vidana, D.~Logoteta, C.~Providencia, A.~Polls, I.~Bombaci, {Estimation of
  the effect of hyperonic three-body forces on the maximum mass of neutron
  stars}, EPL 94~(1) (2011) 11002.
\newblock \href {http://arxiv.org/abs/1006.5660} {\path{arXiv:1006.5660}},
  \href {http://dx.doi.org/10.1209/0295-5075/94/11002}
  {\path{doi:10.1209/0295-5075/94/11002}}.

\bibitem{Yamamoto:2013ada}
Y.~Yamamoto, T.~Furumoto, N.~Yasutake, T.~A. Rijken, {Multi-pomeron repulsion
  and the Neutron-star mass}, Phys. Rev. C 88~(2) (2013) 022801.
\newblock \href {http://arxiv.org/abs/1308.2130} {\path{arXiv:1308.2130}},
  \href {http://dx.doi.org/10.1103/PhysRevC.88.022801}
  {\path{doi:10.1103/PhysRevC.88.022801}}.

\bibitem{Lonardoni:2014bwa}
D.~Lonardoni, A.~Lovato, S.~Gandolfi, F.~Pederiva, {Hyperon Puzzle: Hints from
  Quantum Monte Carlo Calculations}, Phys. Rev. Lett. 114~(9) (2015) 092301.
\newblock \href {http://arxiv.org/abs/1407.4448} {\path{arXiv:1407.4448}},
  \href {http://dx.doi.org/10.1103/PhysRevLett.114.092301}
  {\path{doi:10.1103/PhysRevLett.114.092301}}.

\bibitem{Togashi:2016fky}
H.~Togashi, E.~Hiyama, Y.~Yamamoto, M.~Takano, {Equation of state for neutron
  stars with hyperons by the variational method}, Phys. Rev. C 93~(3) (2016)
  035808.
\newblock \href {http://arxiv.org/abs/1602.08106} {\path{arXiv:1602.08106}},
  \href {http://dx.doi.org/10.1103/PhysRevC.93.035808}
  {\path{doi:10.1103/PhysRevC.93.035808}}.

\bibitem{Weissenborn:2011qu}
S.~Weissenborn, I.~Sagert, G.~Pagliara, M.~Hempel, J.~Schaffner-Bielich, {Quark
  Matter In Massive Neutron Stars}, Astrophys. J. Lett. 740 (2011) L14.
\newblock \href {http://arxiv.org/abs/1102.2869} {\path{arXiv:1102.2869}},
  \href {http://dx.doi.org/10.1088/2041-8205/740/1/L14}
  {\path{doi:10.1088/2041-8205/740/1/L14}}.

\bibitem{Klahn:2013kga}
T.~Kl{\"a}hn, R.~{\L}astowiecki, D.~B. Blaschke, {Implications of the
  measurement of pulsars with two solar masses for quark matter in compact
  stars and heavy-ion collisions: A Nambu{\textendash}Jona-Lasinio model case
  study}, Phys. Rev. D 88~(8) (2013) 085001.
\newblock \href {http://arxiv.org/abs/1307.6996} {\path{arXiv:1307.6996}},
  \href {http://dx.doi.org/10.1103/PhysRevD.88.085001}
  {\path{doi:10.1103/PhysRevD.88.085001}}.

\bibitem{Zhao:2015rta}
T.~Zhao, S.-S. Xu, Y.~Yan, X.-L. Luo, X.-J. Liu, H.-S. Zong, {Studies of
  two-solar-mass hybrid stars within the framework of Dyson-Schwinger
  equations}, Phys. Rev. D 92~(5) (2015) 054012.
\newblock \href {http://arxiv.org/abs/1509.03377} {\path{arXiv:1509.03377}},
  \href {http://dx.doi.org/10.1103/PhysRevD.92.054012}
  {\path{doi:10.1103/PhysRevD.92.054012}}.

\bibitem{Kojo:2014rca}
T.~Kojo, P.~D. Powell, Y.~Song, G.~Baym, {Phenomenological QCD equation of
  state for massive neutron stars}, Phys. Rev. D 91~(4) (2015) 045003.
\newblock \href {http://arxiv.org/abs/1412.1108} {\path{arXiv:1412.1108}},
  \href {http://dx.doi.org/10.1103/PhysRevD.91.045003}
  {\path{doi:10.1103/PhysRevD.91.045003}}.

\bibitem{Masuda:2015kha}
K.~Masuda, T.~Hatsuda, T.~Takatsuka, {Hyperon Puzzle, Hadron-Quark Crossover
  and Massive Neutron Stars}, Eur. Phys. J. A 52~(3) (2016) 65.
\newblock \href {http://arxiv.org/abs/1508.04861} {\path{arXiv:1508.04861}},
  \href {http://dx.doi.org/10.1140/epja/i2016-16065-6}
  {\path{doi:10.1140/epja/i2016-16065-6}}.

\bibitem{Li:2015ida}
A.~Li, W.~Zuo, G.~X. Peng, {Massive hybrid stars with a first order phase
  transition}, Phys. Rev. C 91~(3) (2015) 035803.
\newblock \href {http://arxiv.org/abs/1503.02739} {\path{arXiv:1503.02739}},
  \href {http://dx.doi.org/10.1103/PhysRevC.91.035803}
  {\path{doi:10.1103/PhysRevC.91.035803}}.

\bibitem{Whittenbury:2015ziz}
D.~L. Whittenbury, H.~H. Matevosyan, A.~W. Thomas, {Hybrid stars using the
  quark-meson coupling and proper-time Nambu--Jona-Lasinio models}, Phys. Rev.
  C 93~(3) (2016) 035807.
\newblock \href {http://arxiv.org/abs/1511.08561} {\path{arXiv:1511.08561}},
  \href {http://dx.doi.org/10.1103/PhysRevC.93.035807}
  {\path{doi:10.1103/PhysRevC.93.035807}}.

\bibitem{Fukushima:2015bda}
K.~Fukushima, T.~Kojo, {The Quarkyonic Star}, Astrophys. J. 817~(2) (2016) 180.
\newblock \href {http://arxiv.org/abs/1509.00356} {\path{arXiv:1509.00356}},
  \href {http://dx.doi.org/10.3847/0004-637X/817/2/180}
  {\path{doi:10.3847/0004-637X/817/2/180}}.

\bibitem{Gerstung:2020ktv}
D.~Gerstung, N.~Kaiser, W.~Weise, {Hyperon{\textendash}nucleon three-body
  forces and strangeness in neutron stars}, Eur. Phys. J. A 56~(6) (2020) 175.
\newblock \href {http://arxiv.org/abs/2001.10563} {\path{arXiv:2001.10563}},
  \href {http://dx.doi.org/10.1140/epja/s10050-020-00180-2}
  {\path{doi:10.1140/epja/s10050-020-00180-2}}.

\bibitem{Xie:2019sqb}
W.-J. Xie, B.-A. Li, {Bayesian Inference of High-density Nuclear Symmetry
  Energy from Radii of Canonical Neutron Stars}, Astrophys. J. 883 (2019) 174.
\newblock \href {http://arxiv.org/abs/1907.10741} {\path{arXiv:1907.10741}},
  \href {http://dx.doi.org/10.3847/1538-4357/ab3f37}
  {\path{doi:10.3847/1538-4357/ab3f37}}.

\bibitem{Xie:2020tdo}
W.-J. Xie, B.-A. Li, {Bayesian Inference of the Symmetry Energy of Superdense
  Neutron-rich Matter from Future Radius Measurements of Massive Neutron
  Stars}, Astrophys. J. 899~(1) (2020) 4.
\newblock \href {http://arxiv.org/abs/2005.07216} {\path{arXiv:2005.07216}},
  \href {http://dx.doi.org/10.3847/1538-4357/aba271}
  {\path{doi:10.3847/1538-4357/aba271}}.

\bibitem{Xie:2023llh}
W.-J. Xie, Z.-W. Ma, J.-H. Guo, {Bayesian inference of the crust--core
  transition density via the neutron-star radius and neutron-skin thickness
  data}, Nucl. Sci. Tech. 34~(6) (2023) 91.
\newblock \href {http://dx.doi.org/10.1007/s41365-023-01239-7}
  {\path{doi:10.1007/s41365-023-01239-7}}.

\bibitem{Li:2024imk}
B.-A. Li, X.~Grundler, W.-J. Xie, N.-B. Zhang, {Bayesian inference of fine
  features of the nuclear equation of state from future neutron star radius
  measurements to 0.1~km accuracy}, Phys. Rev. D 110~(10) (2024) 103040.
\newblock \href {http://arxiv.org/abs/2407.07823} {\path{arXiv:2407.07823}},
  \href {http://dx.doi.org/10.1103/PhysRevD.110.103040}
  {\path{doi:10.1103/PhysRevD.110.103040}}.

\bibitem{Malik:2022zol}
T.~Malik, M.~Ferreira, B.~K. Agrawal, C.~Provid{\^e}ncia, {Relativistic
  Description of Dense Matter Equation of State and Compatibility with Neutron
  Star Observables: A Bayesian Approach}, Astrophys. J. 930~(1) (2022) 17.
\newblock \href {http://arxiv.org/abs/2201.12552} {\path{arXiv:2201.12552}},
  \href {http://dx.doi.org/10.3847/1538-4357/ac5d3c}
  {\path{doi:10.3847/1538-4357/ac5d3c}}.

\bibitem{Malik:2022ilb}
T.~Malik, B.~K. Agrawal, C.~Provid{\^e}ncia, {Inferring the nuclear symmetry
  energy at suprasaturation density from neutrino cooling}, Phys. Rev. C
  106~(4) (2022) L042801.
\newblock \href {http://arxiv.org/abs/2206.15404} {\path{arXiv:2206.15404}},
  \href {http://dx.doi.org/10.1103/PhysRevC.106.L042801}
  {\path{doi:10.1103/PhysRevC.106.L042801}}.

\bibitem{Huang:2023grj}
C.~Huang, G.~Raaijmakers, A.~L. Watts, L.~Tolos, C.~Provid{\^e}ncia,
  {Constraining a relativistic mean field model using neutron star
  mass{\textendash}radius measurements I: nucleonic models}, Mon. Not. Roy.
  Astron. Soc. 529~(4) (2024) 4650--4665.
\newblock \href {http://arxiv.org/abs/2303.17518} {\path{arXiv:2303.17518}},
  \href {http://dx.doi.org/10.1093/mnras/stae844}
  {\path{doi:10.1093/mnras/stae844}}.

\bibitem{Li:2024pts}
J.-J. Li, Y.~Tian, A.~Sedrakian, {Bayesian constraints on covariant density
  functional equations of state of compact stars with new NICER mass-radius
  measurements}, Phys. Lett. B 865 (2025) 139501.
\newblock \href {http://arxiv.org/abs/2412.16513} {\path{arXiv:2412.16513}},
  \href {http://dx.doi.org/10.1016/j.physletb.2025.139501}
  {\path{doi:10.1016/j.physletb.2025.139501}}.

\bibitem{Dover1993_NPA560-559}
C. B. Dover and A. Gal, {A generalized Bethe-Weizs\"acker mass formula for strange hadronic matter}, Nucl. Phys. A
  560~(1993) 559.
\newblock \href {https://www.sciencedirect.com/science/article/pii/037594749390115E}
  {\path{doi:10.1016/0375-9474(93)90115-E}}.


\bibitem{Margueron:2017eqc}
J.~Margueron, R.~Hoffmann~Casali, F.~Gulminelli, {Equation of state for dense
  nucleonic matter from metamodeling. I. Foundational aspects}, Phys. Rev. C
  97~(2) (2018) 025805.
\newblock \href {http://arxiv.org/abs/1708.06894} {\path{arXiv:1708.06894}},
  \href {http://dx.doi.org/10.1103/PhysRevC.97.025805}
  {\path{doi:10.1103/PhysRevC.97.025805}}.

\bibitem{Cai:2021ucx}
B.-J. Cai, B.-A. Li, {Auxiliary function approach for determining symmetry
  energy at suprasaturation densities}, Phys. Rev. C 103~(5) (2021) 054611.
\newblock \href {http://arxiv.org/abs/2104.02185} {\path{arXiv:2104.02185}},
  \href {http://dx.doi.org/10.1103/PhysRevC.103.054611}
  {\path{doi:10.1103/PhysRevC.103.054611}}.

\bibitem{Huth:2021bsp}
S.~Huth, et~al., {Constraining Neutron-Star Matter with Microscopic and
  Macroscopic Collisions}, Nature 606 (2022) 276--280.
\newblock \href {http://arxiv.org/abs/2107.06229} {\path{arXiv:2107.06229}},
  \href {http://dx.doi.org/10.1038/s41586-022-04750-w}
  {\path{doi:10.1038/s41586-022-04750-w}}.

\bibitem{Xia:2022dvw}
C.-J. Xia, T.~Maruyama, A.~Li, B.~Y. Sun, W.-H. Long, Y.-X. Zhang, {Unified
  neutron star EOSs and neutron star structures in RMF models}, Commun. Theor.
  Phys. 74 (2022) 095303.
\newblock \href {http://arxiv.org/abs/2208.12893} {\path{arXiv:2208.12893}},
  \href {http://dx.doi.org/10.1088/1572-9494/ac71fd}
  {\path{doi:10.1088/1572-9494/ac71fd}}.

\bibitem{Shlomo:2006ole}
S.~Shlomo, V.~M. Kolomietz, G.~Col{\`o}, {Deducing the nuclear-matter
  incompressibility coefficient from data on isoscalar compression modes}, Eur.
  Phys. J. A 30~(1) (2006) 23--30.
\newblock \href {http://dx.doi.org/10.1140/epja/i2006-10100-3}
  {\path{doi:10.1140/epja/i2006-10100-3}}.

\bibitem{Xie:2020kta}
W.-J. Xie, B.-A. Li, {Bayesian inference of the incompressibility, skewness and
  kurtosis of nuclear matter from empirical pressures in relativistic heavy-ion
  collisions}, J. Phys. G 48~(2) (2021) 025110.
\newblock \href {http://arxiv.org/abs/2001.03669} {\path{arXiv:2001.03669}},
  \href {http://dx.doi.org/10.1088/1361-6471/abd25a}
  {\path{doi:10.1088/1361-6471/abd25a}}.

\bibitem{Tsang:2008fd}
M.~B. Tsang, Y.~Zhang, P.~Danielewicz, M.~Famiano, Z.~Li, W.~G. Lynch, A.~W.
  Steiner, {Constraints on the density dependence of the symmetry energy},
  Phys. Rev. Lett. 102 (2009) 122701.
\newblock \href {http://arxiv.org/abs/0811.3107} {\path{arXiv:0811.3107}},
  \href {http://dx.doi.org/10.1103/PhysRevLett.102.122701}
  {\path{doi:10.1103/PhysRevLett.102.122701}}.

\bibitem{Zhang:2020azr}
Y.~Zhang, M.~Liu, C.-J. Xia, Z.~Li, S.~K. Biswal, {Constraints on the symmetry
  energy and its associated parameters from nuclei to neutron stars}, Phys.
  Rev. C 101~(3) (2020) 034303.
\newblock \href {http://arxiv.org/abs/2002.10884} {\path{arXiv:2002.10884}},
  \href {http://dx.doi.org/10.1103/PhysRevC.101.034303}
  {\path{doi:10.1103/PhysRevC.101.034303}}.

\bibitem{Prakash:1992zng}
M.~Prakash, M.~Prakash, J.~M. Lattimer, C.~J. Pethick, {Rapid cooling of
  neutron stars by hyperons and Delta isobars}, Astrophys. J. Lett. 390 (1992)
  L77.
\newblock \href {http://dx.doi.org/10.1086/186376} {\path{doi:10.1086/186376}}.

\bibitem{Lattimer:1991ib}
J.~M. Lattimer, M.~Prakash, C.~J. Pethick, P.~Haensel, {Direct URCA process in
  neutron stars}, Phys. Rev. Lett. 66 (1991) 2701--2704.
\newblock \href {http://dx.doi.org/10.1103/PhysRevLett.66.2701}
  {\path{doi:10.1103/PhysRevLett.66.2701}}.

\bibitem{Guan:2025yia}
Z.~Y. Guan, Y.~F. Niu, {Reconciliation between Neutron Skin Thickness from
  PREX-2 Experiment and Neutron-Star Tidal Polarizability from GW170817 Event:
  The Key Role of Symmetry Energy Curvature}, Phys. Rev. Lett. 135~(17) (2025)
  172701.
\newblock \href {http://dx.doi.org/10.1103/6m4p-wmv3}
  {\path{doi:10.1103/6m4p-wmv3}}.

\bibitem{Doroshenko:2022nwp}
V.~Doroshenko, V.~Suleimanov, G.~P{\"u}hlhofer, A.~Santangelo, {A strangely
  light neutron star within a supernova remnant}, Nature Astron. 6~(12) (2022)
  1444--1451.
\newblock \href {http://dx.doi.org/10.1038/s41550-022-01800-1}
  {\path{doi:10.1038/s41550-022-01800-1}}.

\bibitem{Mauviard:2025dmd}
L.~Mauviard, et~al., {A NICER View of the 1.4 M$_{\odot}$ Edge-on Pulsar PSR
  J0614-3329}, Astrophys. J. 995~(1) (2025) 60.
\newblock \href {http://arxiv.org/abs/2506.14883} {\path{arXiv:2506.14883}},
  \href {http://dx.doi.org/10.3847/1538-4357/ae145d}
  {\path{doi:10.3847/1538-4357/ae145d}}.

\bibitem{Riley:2019yda}
T.~E. Riley, et~al., {A $NICER$ View of PSR J0030+0451: Millisecond Pulsar
  Parameter Estimation}, Astrophys. J. Lett. 887~(1) (2019) L21.
\newblock \href {http://arxiv.org/abs/1912.05702} {\path{arXiv:1912.05702}},
  \href {http://dx.doi.org/10.3847/2041-8213/ab481c}
  {\path{doi:10.3847/2041-8213/ab481c}}.

\bibitem{Choudhury:2024xbk}
D.~Choudhury, et~al., {A NICER View of the Nearest and Brightest Millisecond
  Pulsar: PSR J0437--4715}, Astrophys. J. Lett. 971~(1) (2024) L20.
\newblock \href {http://arxiv.org/abs/2407.06789} {\path{arXiv:2407.06789}},
  \href {http://dx.doi.org/10.3847/2041-8213/ad5a6f}
  {\path{doi:10.3847/2041-8213/ad5a6f}}.

\bibitem{LIGOScientific:2018cki}
B.~P. Abbott, et~al., {GW170817: Measurements of neutron star radii and
  equation of state}, Phys. Rev. Lett. 121~(16) (2018) 161101.
\newblock \href {http://arxiv.org/abs/1805.11581} {\path{arXiv:1805.11581}},
  \href {http://dx.doi.org/10.1103/PhysRevLett.121.161101}
  {\path{doi:10.1103/PhysRevLett.121.161101}}.

\bibitem{Annala:2023cwx}
E.~Annala, T.~Gorda, J.~Hirvonen, O.~Komoltsev, A.~Kurkela, J.~N{\"a}ttil{\"a},
  A.~Vuorinen, {Strongly interacting matter exhibits deconfined behavior in
  massive neutron stars}, Nature Commun. 14~(1) (2023) 8451.
\newblock \href {http://arxiv.org/abs/2303.11356} {\path{arXiv:2303.11356}},
  \href {http://dx.doi.org/10.1038/s41467-023-44051-y}
  {\path{doi:10.1038/s41467-023-44051-y}}.

\bibitem{Zhao:2022toc}
T.~Zhao, C.~Constantinou, P.~Jaikumar, M.~Prakash, {Quasinormal g modes of
  neutron stars with quarks}, Phys. Rev. D 105~(10) (2022) 103025.
\newblock \href {http://arxiv.org/abs/2202.01403} {\path{arXiv:2202.01403}},
  \href {http://dx.doi.org/10.1103/PhysRevD.105.103025}
  {\path{doi:10.1103/PhysRevD.105.103025}}.

\bibitem{Zhang:2025rnf}
C.~Zhang, J.~M.~Z. Pretel, R.~Xu, {Self-bound hybrid stars with strong phase
  transitions can relieve major compact star observation tensions}, Phys. Rev.
  D 113~(4) (2026) 043006.
\newblock \href {http://arxiv.org/abs/2507.01371} {\path{arXiv:2507.01371}},
  \href {http://dx.doi.org/10.1103/q9pg-9ch5} {\path{doi:10.1103/q9pg-9ch5}}.

\bibitem{Yuan:2024hge}
W.-L. Yuan, C.~Huang, C.~Zhang, E.~Zhou, R.~Xu, {Bayesian inference of
  strangeon matter using the measurements of PSR J0437-4715 and GW190814},
  Phys. Rev. D 111~(6) (2025) 063033.
\newblock \href {http://arxiv.org/abs/2411.14938} {\path{arXiv:2411.14938}},
  \href {http://dx.doi.org/10.1103/PhysRevD.111.063033}
  {\path{doi:10.1103/PhysRevD.111.063033}}.

\bibitem{Miller:2021qha}
M.~C. Miller, et~al., {The Radius of PSR J0740+6620 from NICER and XMM-Newton
  Data}, Astrophys. J. Lett. 918~(2) (2021) L28.
\newblock \href {http://arxiv.org/abs/2105.06979} {\path{arXiv:2105.06979}},
  \href {http://dx.doi.org/10.3847/2041-8213/ac089b}
  {\path{doi:10.3847/2041-8213/ac089b}}.

\bibitem{Huang:2024rfg}
C.~Huang, et~al., {CompactObject: An open-source Python package for full-scope
  neutron star equation of state inference}\href
  {http://arxiv.org/abs/2411.14615} {\path{arXiv:2411.14615}}.

\bibitem{Drischler:2020hwi}
C.~Drischler, R.~J. Furnstahl, J.~A. Melendez, D.~R. Phillips, {How Well Do We
  Know the Neutron-Matter Equation of State at the Densities Inside Neutron
  Stars? A Bayesian Approach with Correlated Uncertainties}, Phys. Rev. Lett.
  125~(20) (2020) 202702.
\newblock \href {http://arxiv.org/abs/2004.07232} {\path{arXiv:2004.07232}},
  \href {http://dx.doi.org/10.1103/PhysRevLett.125.202702}
  {\path{doi:10.1103/PhysRevLett.125.202702}}.

\bibitem{Drischler:2020yad}
C.~Drischler, J.~A. Melendez, R.~J. Furnstahl, D.~R. Phillips, {Quantifying
  uncertainties and correlations in the nuclear-matter equation of state},
  Phys. Rev. C 102~(5) (2020) 054315.
\newblock \href {http://arxiv.org/abs/2004.07805} {\path{arXiv:2004.07805}},
  \href {http://dx.doi.org/10.1103/PhysRevC.102.054315}
  {\path{doi:10.1103/PhysRevC.102.054315}}.

\bibitem{Drischler:2017wtt}
C.~Drischler, K.~Hebeler, A.~Schwenk, {Chiral interactions up to
  next-to-next-to-next-to-leading order and nuclear saturation}, Phys. Rev.
  Lett. 122~(4) (2019) 042501.
\newblock \href {http://arxiv.org/abs/1710.08220} {\path{arXiv:1710.08220}},
  \href {http://dx.doi.org/10.1103/PhysRevLett.122.042501}
  {\path{doi:10.1103/PhysRevLett.122.042501}}.

\bibitem{Fuchs:2005zg}
C.~Fuchs, {Kaon production in heavy ion reactions at intermediate energies},
  Prog. Part. Nucl. Phys. 56 (2006) 1--103.
\newblock \href {http://arxiv.org/abs/nucl-th/0507017}
  {\path{arXiv:nucl-th/0507017}}, \href
  {http://dx.doi.org/10.1016/j.ppnp.2005.07.004}
  {\path{doi:10.1016/j.ppnp.2005.07.004}}.

\bibitem{Lynch:2009vc}
W.~G. Lynch, M.~B. Tsang, Y.~Zhang, P.~Danielewicz, M.~Famiano, Z.~Li, A.~W.
  Steiner, {Probing the Symmetry Energy with Heavy Ions}, Prog. Part. Nucl.
  Phys. 62 (2009) 427.
\newblock \href {http://arxiv.org/abs/0901.0412} {\path{arXiv:0901.0412}},
  \href {http://dx.doi.org/10.1016/j.ppnp.2009.01.001}
  {\path{doi:10.1016/j.ppnp.2009.01.001}}.

\bibitem{Danielewicz:2002pu}
P.~Danielewicz, R.~Lacey, W.~G. Lynch, {Determination of the equation of state
  of dense matter}, Science 298 (2002) 1592--1596.
\newblock \href {http://arxiv.org/abs/nucl-th/0208016}
  {\path{arXiv:nucl-th/0208016}}, \href
  {http://dx.doi.org/10.1126/science.1078070}
  {\path{doi:10.1126/science.1078070}}.

\bibitem{Riley:2021pdl}
T.~E. Riley, et~al., {A NICER View of the Massive Pulsar PSR J0740+6620
  Informed by Radio Timing and XMM-Newton Spectroscopy}, Astrophys. J. Lett.
  918~(2) (2021) L27.
\newblock \href {http://arxiv.org/abs/2105.06980} {\path{arXiv:2105.06980}},
  \href {http://dx.doi.org/10.3847/2041-8213/ac0a81}
  {\path{doi:10.3847/2041-8213/ac0a81}}.

\bibitem{Miller:2019cac}
M.~C. Miller, et~al., {PSR J0030+0451 Mass and Radius from $NICER$ Data and
  Implications for the Properties of Neutron Star Matter}, Astrophys. J. Lett.
  887~(1) (2019) L24.
\newblock \href {http://arxiv.org/abs/1912.05705} {\path{arXiv:1912.05705}},
  \href {http://dx.doi.org/10.3847/2041-8213/ab50c5}
  {\path{doi:10.3847/2041-8213/ab50c5}}.

\bibitem{Greif:2018njt}
S.~K. Greif, G.~Raaijmakers, K.~Hebeler, A.~Schwenk, A.~L. Watts, {Equation of
  state sensitivities when inferring neutron star and dense matter properties},
  Mon. Not. Roy. Astron. Soc. 485~(4) (2019) 5363--5376.
\newblock \href {http://arxiv.org/abs/1812.08188} {\path{arXiv:1812.08188}},
  \href {http://dx.doi.org/10.1093/mnras/stz654}
  {\path{doi:10.1093/mnras/stz654}}.

\bibitem{Raaijmakers:2019qny}
G.~Raaijmakers, et~al., {A $NICER$ view of PSR J0030+0451: Implications for the
  dense matter equation of state}, Astrophys. J. Lett. 887~(1) (2019) L22.
\newblock \href {http://arxiv.org/abs/1912.05703} {\path{arXiv:1912.05703}},
  \href {http://dx.doi.org/10.3847/2041-8213/ab451a}
  {\path{doi:10.3847/2041-8213/ab451a}}.

\bibitem{Raaijmakers:2019dks}
G.~Raaijmakers, et~al., {Constraining the dense matter equation of state with
  joint analysis of NICER and LIGO/Virgo measurements}, Astrophys. J. Lett.
  893~(1) (2020) L21.
\newblock \href {http://arxiv.org/abs/1912.11031} {\path{arXiv:1912.11031}},
  \href {http://dx.doi.org/10.3847/2041-8213/ab822f}
  {\path{doi:10.3847/2041-8213/ab822f}}.

\bibitem{Foreman-Mackey:2012any}
D.~Foreman-Mackey, D.~W. Hogg, D.~Lang, J.~Goodman, {emcee: The MCMC Hammer},
  Publ. Astron. Soc. Pac. 125 (2013) 306--312.
\newblock \href {http://arxiv.org/abs/1202.3665} {\path{arXiv:1202.3665}},
  \href {http://dx.doi.org/10.1086/670067} {\path{doi:10.1086/670067}}.


\bibitem{Ahn2013_PRC88-014003}
J. K. Ahn, H. Akikawa, S. Aoki, K. Arai, S. Y. Bahk, K. M. Baik, B. Bassalleck, J. H. Chung et al. (E373 (KEK-PS) Collaboration), Phys. Rev. C 88 (2013) 014003.
\newblock \href {http://dx.doi.org/10.1103/PhysRevC.88.014003} {\path{doi:10.1103/PhysRevC.88.014003}}.


\end{thebibliography}
\end{document}